\documentclass[twocolumn,prd,superscriptaddress,nofootinbib]{revtex4}

\usepackage{amsmath,color,graphicx}
\usepackage[colorlinks=true,citecolor=blue,urlcolor=blue]{hyperref}

\newcommand{\SI}[2]{\ensuremath{#1\,{\rm #2}}}

\newcommand{\specialcell}[2][c]{%
  \begin{tabular}[#1]{@{}c@{}}#2\end{tabular}}

\begin{document}
\title{Mitigation of the instrumental noise transient in gravitational-wave data surrounding GW170817}

\author{Chris Pankow}
\affiliation{Center for Interdisciplinary Research and Exploration in Astrophysics (CIERA) and Department of Physics and Astronomy, Northwestern University, 2145 Sheridan Road, Evanston, IL 60208, USA}

\author{Katerina Chatziioannou}
\affiliation{Canadian Institute for Theoretical Astrophysics, 60 St. George Street, Toronto, Ontario, M5S 3H8, Canada}

\author{Eve A. Chase}
\affiliation{Center for Interdisciplinary Research and Exploration in Astrophysics (CIERA) and Department of Physics and Astronomy, Northwestern University, 2145 Sheridan Road, Evanston, IL 60208, USA}

\author{Tyson B. Littenberg}
\affiliation{NASA Marshall Space Flight Center, Huntsville, Alabama 35811, USA}

\author{Matthew Evans}
\affiliation{LIGO, Massachusetts Institute of Technology, Cambridge, Massachusetts 02139, USA}

\author{Jessica McIver}
\affiliation{LIGO, California Institute of Technology, Pasadena, CA 91125, USA}

\author{Neil J. Cornish}
\affiliation{eXtreme Gravity Institute, Department of Physics, Montana State University, Bozeman, Montana 59717, USA}

\author{Carl-Johan Haster}
\affiliation{Canadian Institute for Theoretical Astrophysics, 60 St. George Street, Toronto, Ontario, M5S 3H8, Canada}

\author{Jonah Kanner}
\affiliation{LIGO, California Institute of Technology, Pasadena, CA 91125, USA}

\author{Vivien Raymond}
\affiliation{Cardiff University, Cardiff CF24 3AA, United Kingdom}

\author{Salvatore Vitale}
\affiliation{LIGO Laboratory and Kavli Institute for Astrophysics and Space Research, Massachusetts Institute of Technology, Cambridge, Massachusetts 02139, USA }

\author{Aaron Zimmerman}
\affiliation{Canadian Institute for Theoretical Astrophysics, 60 St. George Street, Toronto, Ontario, M5S 3H8, Canada}

\begin{abstract}
In the coming years gravitational-wave detectors will undergo a series of improvements, with an increase in their detection rate by about an order of magnitude. Routine detections of gravitational-wave signals promote novel astrophysical and fundamental theory studies, while simultaneously leading to an increase in the number of detections temporally overlapping with instrumentally- or environmentally-induced transients in the detectors (glitches), often of unknown origin. Indeed, this was the case for the very first detection by the LIGO and Virgo detectors of a gravitational-wave signal consistent with a binary neutron star coalescence, GW170817. A loud glitch in the LIGO-Livingston detector, about one second before the merger, hampered coincident detection (which was initially achieved solely with LIGO-Hanford data). Moreover, accurate source characterization depends on specific assumptions about the behavior of the detector noise that are rendered invalid by the presence of glitches. In this paper, we present the various techniques employed for the initial mitigation of the glitch to perform source characterization of GW170817 and study advantages and disadvantages of each mitigation method. We show that, despite the presence of instrumental noise transients louder than the one affecting GW170817, we are still able to produce unbiased measurements of the intrinsic parameters from simulated injections with properties similar to GW170817.
\end{abstract}

\maketitle

\section{Introduction}
\label{sec:intro}

The discovery of GW170817~\citep{TheLIGOScientific:2017qsa} was a watershed moment in astrophysics. It was the first detection of a binary neutron star (BNS) through gravitational waves (GWs), and its association with GRB170817A~\citep{Monitor:2017mdv} spurred a world-wide effort which revealed emission throughout the electromagnetic spectrum~\citep{GBM:2017lvd}.
The signal is the loudest GW event detected so far, exhibiting a signal-to-noise ratio (SNR) of $18.8$ in the LIGO-Hanford, $26.4$ in the LIGO-Livingston~\cite{TheLIGOScientific:2014jea}, and $2$ in the Virgo~\cite{TheVirgo:2014hva} detectors. 

Initially identified by low-latency compact binary searches~\citep{PhysRevD.95.042001,2018arXiv180511174N}, coincidence and rapid localization~\citep{PhysRevD.93.024013} were hindered because of a non-astrophysical transient present in the LIGO-Livingston detector about one second before the estimated merger time of the signal~\citep{TheLIGOScientific:2017qsa}. 
This transient signal (colloquially referred to as a \textit{glitch}~\citep{2015CQGra..32x5005N,2016CQGra..33m4001A}) was a high-amplitude, short-duration excursion of unknown origin in the control loop used to sense the differential motion of the interferometer arms, which is calibrated to produce GW strain data~\citep{TheLIGOScientific:2014jea}. 
This excursion caused an overflow in the digital-to-analog converter of the optic drive signal and prevented the searches from registering an event trigger for GW170817 in the LIGO-Livingston detector in low latency. 
The cause of the excursion is unknown, but such \emph{overflow} glitches occur independently at a rate of roughly once every few hours in both LIGO detectors~\citep{2018CQGra..35f5010A}.

After the signal was visually identified in both LIGO detectors, it was clear that the noise transient in LIGO-Livingston would need to be mitigated in order to provide accurate sky localization maps and estimate the parameters of GW170817.

Three types of glitch removal were employed.  The first method excised 0.2 seconds of time series data corrupted by the noise transient using the PyCBC \emph{gating} algorithm~\cite{2016CQGra..33u5004U,pycbc-software}. The second method is a variation on the first which allows for better fidelity in the process. 
The gating methods were developed in the context of detection, since glitches can corrupt the output of matched filtering searches~\citep{2016CQGra..33u5004U,PhysRevD.95.042001}. They are applied in an automated processes which identify statistically significant excursions in the data stream and excise them.
Cognizant that any electromagnetic counterpart was likely to fade on the order of hours to days~\citep{2012ApJ...746...48M}, the first method of data excision was used to produce and promptly disseminate an accurate sky map for GW170817~\cite{GCN21513}. 
This resulted in the successful discovery of a counterpart electromagnetic signal~\citep{2017CoulterNat}.
However, while such data excision techniques are speedy to apply, portions of the underlying signal are also removed in the process. 
A third, more sophisticated, technique used the {\tt BayesWave} algorithm~\citep{0264-9381-32-13-135012} to model the glitch and subtract it from the data, preserving more of the signal. 
The glitch-subtracted data, produced days after the initial detection, was used for parameter estimation~\citep{TheLIGOScientific:2017qsa} and more detailed studies of the signal properties~\citep{TheLIGOScientific:2018pe, TheLIGOScientific:2018eos}. 

The focus of this work is to evaluate the efficacy and fidelity of these techniques --- \emph{gating} and {\tt BayesWave} --- in the context of characterization of the intrinsic properties of the signal. A detailed study of the effect of gating and {\tt BayesWave} cleaning on sky position reconstruction is deferred to a later study.
In our analysis we use the  {\tt LALInference} package~\cite{Veitch:2014wba}, a standard parameter estimation analysis for GW signals from compact binaries. 
This analysis models the signal with existing theoretical waveform templates and assumes that the noise is \emph{stationary}, namely that its properties do not change appreciably on the timescale of the signal, and \emph{Gaussian}, namely that the noise follows a colored Gaussian distribution. The presence of the noise transient in the data breaks both assumptions about noise behavior.

For a true GW source, such as GW170817, no control case is available for a glitch mitigation study; the true waveform and intrinsic source parameters are not known \emph{a priori}. We circumvent this problem with a set of synthetic event injection studies, using data stretches which exhibit instrumental transients similar to the one afflicting GW170817. We inject synthetic GW170817-like signals concurrently with those noise transients and then apply the transient removal methods. We then recover the signal parameters, and compare the known properties of the injected synthetic signal to the posterior distributions obtained by analyzing the data with and without the subtraction procedure applied. As a reference, we also perform the same parameter estimation analysis with a zero-noise realization. This provides an ideal case for comparison with the real-noise analyses.

We find that modeling and subtracting the glitch using the {\tt BayesWave} algorithm leads to robust and unbiased estimation of the parameters of the signal, including the masses, spins, and tidal deformabilities of the binary components. The latter is particularly relevant in the case of GW170817. The tidal deformability quantifies the deformation of a body in the presence of an external field~\citep{flanagan:021502} and depends sensitively on the equation of state of matter at supranuclear densities~\citep{2008ApJ...677.1216H,2010PhRvD..81l3016H}. Since it affects the binary evolution mostly during the latest stages~\citep{Read:2009} of its evolution, its main effect on the waveform temporally coincides with the glitch in the data (which, as mentioned above, happened roughly 1 second before the estimated merger time). Despite this, we show that the measurement of the tidal deformability presented in Refs.~\citep{TheLIGOScientific:2017qsa, TheLIGOScientific:2018pe, TheLIGOScientific:2018eos} are only minimally affected by the glitch once it has been successfully modeled and removed.

\section{Mitigation techniques}
\label{sec:glitch_removal}

In the presence of a non-astrophysical transient and a GW signal, the measured strain time series can be expressed~\citep{1993PhRvD..47.2198F,1998PhRvD..57.4535F} as
\begin{equation}
\label{eqn:strain}
s(t) = n(t) + h(t, \vec{\lambda}) + g(t, \Delta t)\,.
\end{equation}
In the above equation $n(t)$ is a stream of stationary and Gaussian noise, $h(t)$ is the GW signal that depends on the binary parameters $\vec{\lambda}$~\citep{2012PhRvD..85l2006A}, and $g(t)$ is the non-Gaussian, instrumentally-induced transient of duration $\Delta t$. The spectral profile of the noise is characterized by the one-sided power spectral density (PSD) denoted $S(f)$. The detection and parameter estimation algorithms make explicit assumptions about the nature of the data --- namely that the Fourier transform of $n(t)$, $\tilde{n}(f)$ is related to the PSD by $2\langle \tilde{n}(f) \tilde{n}^{*}(f') \rangle = S(f)\delta(f-f')$. In other words, the noise is stationary and Gaussian distributed, given its ensemble average in the Fourier domain is proportional to the spectral density~\cite{Sathyaprakash:2009xs}.

In the context of GW parameter estimation both of these assumptions determine the form of the log-likelihood function for the data, given a model for the GW signal $h(\vec{\lambda}')$~\cite{Veitch:2014wba}, defined as:
\begin{equation}
\log L(s|\vec{\lambda}')\propto \langle\tilde{s} - \tilde{h}(\vec{\lambda}')|\tilde{s} - \tilde{h}(\vec{\lambda}')\rangle\label{eqn:likelihood}.
\end{equation}
where the quantities with a tilde are the Fourier-domain transformed quantities from the time-domain equivalents in Eq.~\eqref{eqn:strain}. The brackets denote a noise-weighted inner-product defined as $\langle a|b\rangle \equiv 4 \Re \int_0^{\infty} \tilde{a}^{*}\tilde{b} / S(f) df$.

Eq.~\eqref{eqn:likelihood} is calculated from the residual after subtracting the expected signal $\tilde{h}(\vec{\lambda}')$ from the data $\tilde{s}$. 
In the presence of a glitch, the residual consists of both Gaussian noise and the instrumentally-induced transient. 
Moreover, the parameters of the signal can be severely biased as the analysis attempts to return residuals that are consistent with our assumption of Gaussian noise.
The intrinsic amplitude of the glitch relative to the Gaussian noise sets the scale of the mismatched residual (e.g. with which $h$ can $g$ be matched), and if the glitch is comparable in SNR relative to the signal, it can overwhelm the likelihood calculated for the signal alone~\citep{Powell:2018csz}. For a broad overview of the impact on the estimation of the signals properties of compact object coalescences, including the effects of longer-duration glitches, see~\citep{McIver2018}. We next describe strategies to mitigate glitch-induced biases by removing the glitch from the data, $s(t)$, with minimal impact of the recovered signal.

\subsection{Frequency-independent subtraction}
\label{sec:tukey_sub}

The initial solution to mitigating the noise transient near GW170817 was to smoothly taper the time series data to zero while the glitch was present~\cite{2016CQGra..33u5004U,pycbc-software}, see Fig.~2 of Ref.~\citep{TheLIGOScientific:2017qsa},
\begin{equation}
\label{eqn:tukey_sub}
W(t) = \left[1 - w(t, t_0, t_w, t_\mathrm{taper})\right]\,.
\end{equation}
In the above a windowing function $w$ --- typically a Tukey window --- is applied around the center of the noise transient time $t_0$. The shape of the window is set by $t_w$, the half-duration of the data to zero out, and $t_\mathrm{taper}$, the duration of the Tukey tapering on each side of the gated data. For GW170817, $t_0 = 1187008881.389$, $t_w = 0.1s$, and $t_\mathrm{taper} = 0.5s$. The duration of the data affected by the gating window $\Delta t = 2 t_w + 2 t_\mathrm{taper}$~\cite{TheLIGOScientific:2017qsa}. Clearly, this also removes the signal $h(t)$ during $\Delta t$ from the data series. Therefore, while this method is effective in removing the high amplitude power in the glitch, it might induce a different bias because the signal waveform template would also need to be windowed in the same way to match optimally. Section \ref{sec:gw170817_glitch} shows that, for a zeroed time window of $2t_w = 0.25$ seconds or less, this bias is within the tolerances set by the uncertainties in the posterior distributions for a preliminary analysis, and so the trade-off is acceptable.

\subsection{Time-frequency area subtraction}
\label{sec:tf_sub}

A similar, but \emph{frequency-dependent} solution was developed to further reduce damaging the fidelity of the noise-transient-subtracted data $s'(t)$ 

\begin{equation}
\label{eqn:tf_sub}
s'(t) = s(t) - g_{\rm TFA}(t, t_0, \Delta t, \Delta f)\,,
\end{equation}
 where $g_{\rm TFA}(t, t_0, \Delta t, \Delta f)$ is the glitch estimate produced by the Time-Frequency Area (TFA) algorithm.

The TFA algorithm starts by band-passing the data around the noise transient in the
 frequency band effected by the glitch (e.g., $50$ to \SI{800}{Hz}).
This is done in the time-domain after de-trending and windowing the time-series.

The Fast Fourier Transform (FFT) of the band-passed data is then computed,
 and the amplitudes of frequencies corresponding to known strong features 
 (e.g., calibration lines) are set to zero. 
If desired, additional band-passing can also be done in the frequency domain
 by setting the amplitude of out of band frequencies to zero as well.
For the results presented here, frequencies near \SI{300}{Hz} and \SI{500}{Hz}
 were set to zero to remove strong narrow lines due to injected calibration lines and mechanical resonances \citep{2016DetChar.150914}, and frequencies above \SI{700}{Hz}
 were zeroed to improve the band-pass filtering.

Finally, the band-passed and line-removed frequency-domain data are converted back
 to the time domain with an inverse FFT operation, and the glitch estimate
 $g_{\rm TFA}(t, t_0, \Delta t, \Delta f)$ is produced
 by windowing this time-series around the time of the glitch.
The Tukey window function, which is equal to $1$ in the central region,
 is used throughout the TFA algorithm to ensure that the amplitude of the 
 de-glitched time series is not modified by the windowing operation.

\subsection{Noise-transient fitting and removal}
\label{sec:bw_sub}

Apart from generic considerations about the duration and bandwidth of the noise transient, the methods described in Secs.~\ref{sec:tukey_sub} and \ref{sec:tf_sub} are mostly agnostic to the morphology of the transient itself. They are expedient to apply, and significantly reduce biases in the recovered signal parameters. However, the signal is still modified in some way because of the excision applied. A more sophisticated treatment can be applied if one considers the coherent and incoherent decompositions of the data from the two instruments. While a true GW signal is \emph{coherent} across the detector network, any instrumentally-induced transient is \emph{incoherent}, assuming that the noise in the various detectors is uncorrelated. The morphology-independent algorithm {\tt BayesWave}~\citep{0264-9381-32-13-135012} exploits this distinction to reconstruct any coherent and incoherent power in the detectors through a continuous wavelet basis. {\tt BayesWave} uses Morlet-Gabor wavelets, with morphologies
\begin{equation}
\Psi(t; A, f_{0}, \tau, t_{0},\phi_0) = A e^{-(t-t_0)^2/\tau^2} \cos{[2\pi f_0(t\!-\!t_0)\!+\!\phi_0]}
\end{equation}

\noindent to simultaneously model both the coherent signal and the incoherent noise transient in Eq.~\eqref{eqn:strain}. Each wavelet depends on five parameters: its amplitude $A$, quality factor $Q\equiv 2\pi f_0 \tau$, central frequency $f_0$, central time $t_0$, and phase $\phi_0$. 

{\tt BayesWave} harnesses the inherent efficiency of Bayesian inference~\citep{0264-9381-32-13-135012}. 
When a signal is coherent across the detector network it can be fully described in all detectors by the same set of Morlet-Gabor wavelets plus four extrinsic parameters characterizing its sky location, polarization, and the ratio of the two independent GW polarization amplitudes (ellipticity). An incoherent signal, on the other hand, can only be reconstructed with independent sets of wavelets in each detector. Since the incoherent model (termed the `glitch model' in {\tt BayesWave} literature) has more parameters ($ 5N_d N$, where $N_d$ is the number of detectors and $N$ the number of wavelets) than the coherent model ($5N+4$, usually referred to as the `signal model'), the latter will be preferred assuming both models fit the data equally well~\citep{Veitch:2009hd}. Any coherent signal can be separated from the incoherent power, as it can be modeled with fewer parameters~\citep{Kanner2015}. 

In practice and for the signal and glitch near GW170817, this process is aided by the fact that a BNS inspiral and an overflow glitch are reconstructed by wavelets of largely different quality factors, $Q$. A BNS signal is characterized by a long duration (of order seconds to minutes), and is therefore best modeled by wavelets with a large quality factor. On the contrary, the glitch has a short duration and a more compact time-frequency signature, which is best recovered with wavelets with a small quality factor.

Once the coherent and the incoherent part of the signal have been simultaneously reconstructed, the incoherent portion of the data is subtracted from the individual instrument data streams. The instrumental transient --- present only in the LIGO-Livingston detector --- is removed while the coherent GW signal is only minimally affected, giving
\begin{equation}
\label{eqn:bw_sub}
s'(t) = s(t) - g_I(t),
\end{equation}
where $g_I(t)$ is {\tt BayesWave}'s reconstruction of the incoherent part of the data. 
In effect, {\tt BayesWave} fits for the shape of $g(t)$ while leaving $h(t)$ and $n(t)$ unaffected in Eq.~\eqref{eqn:strain}. As a consequence, the assumptions of stationary and Gaussian residuals once the GW signal has been subtracted are explicitly restored and normal parameter estimation techniques are applicable. We show in Sec.~\ref{sec:glitch_pe} that this procedure leads to negligible biases in the parameters of the signal.

\section{Signal characterization}
\label{sec:characterization}

The \textit{overflow} glitch described in Sec.~\ref{sec:intro} preceded the merger time of GW170817 in the LIGO-Livingston instrument by about $1.1$ seconds. As a result, the signal was originally only identified in the LIGO-Hanford detector. In addition to automated monitors~\cite{2013PhRvD..88f2003B,2013CQGra..30o5010E}, manual checks and visual inspections of the data revealed both the glitch in LIGO-Livingston and the coincident nature of the signal. The glitch intersected the time-frequency track of the merger signal, immediately revealing the need to mitigate the impact of the glitch. In order to produce a timely alert of the event to partner astronomers, the frequency-independent data excision method of Sec.~\ref{sec:tukey_sub} was used to generate a rapid sky position posterior. On longer time scales, the methods in \ref{sec:tf_sub} and \ref{sec:bw_sub} were applied, and ultimately, the final results presented in \cite{TheLIGOScientific:2017qsa, TheLIGOScientific:2018pe, TheLIGOScientific:2018eos} used the data set produced with the method in \ref{sec:bw_sub}.

In this section, we compare the performance and fidelity of methods \ref{sec:tukey_sub}, \ref{sec:tf_sub}, and \ref{sec:bw_sub} by injecting synthetic signals with parameters comparable to GW170817 in data from the two LIGO detectors. 
In Sec.~\ref{sec:gw170817_glitch} we compare the effect of glitch-mitigation methods of Secs.~\ref{sec:tukey_sub},~\ref{sec:tf_sub}, and~\ref{sec:bw_sub} on a simulated signal in the absence of a glitch.
Since the data are completely excised, the absence of a glitch is irrelevant to the investigation of the two gating methods.
The use of {\tt BayesWave} glitch-removal method on data where there is no glitch present should meanwhile change the results very little, which we confirm.
We then explore the performance of {\tt BayesWave} glitch removal in the analysis of signals injected on top of actual instrumental glitches in~\ref{sec:glitch_pe}. 
For all the analyses of this section the PSD is calculated from on-source data with the technique presented in~\cite{Littenberg:2014oda}. We also marginalize over the calibration uncertainty of the detector as explained in~\cite{SplineCalMarg-T1400682}. For simplicity, we assume priors on the amplitude (phase) marginalization spline points of $5\%\,(3^\circ)$.

\subsection{Comparison of mitigation methods}
\label{sec:gw170817_glitch}

In this subsection, we study the effect of the three glitch-mitigation methods described in Sec.~\ref{sec:glitch_removal} on source characterization. 
We select a time less than a half-hour after GW170817 in a contiguous set of data and hence preserve the state of the instruments as closely as possible and add a simulated GW170817-like signal to the data, using the {\tt IMRPhenomPv2} waveform family, see Sec.~\ref{sec:glitch_pe}.
For this specific time, no glitch was present within the duration of the simulated signal.
The methods from \ref{sec:tukey_sub}, \ref{sec:tf_sub}, and \ref{sec:bw_sub} were applied to the simulated signal in exactly the same manner and temporal offset relative to GW170817.

In particular, understanding the consequences of the two glitch excision methods (Secs.~\ref{sec:tukey_sub} and~\ref{sec:tf_sub}) is important since the methods are straightforward and quick to apply, and will likely continue to be used for rapid response to interesting GW events, especially as the detection rate increases and the probability of signals overlapping with glitches rises. 
The initial data excision is expected to impact the extracted parameters differently depending on the intersection of the excision and time-frequency track of the signal. In the case of GW170817, the intersection point of the track with the excision is relatively close to the merger time; the fixed time interval in between the glitch peak time and the signal implies the intersection point in the time-frequency plane is near $\sim 100$ Hz. As a result, the sweep of the signal through the most sensitive part of the instrument bandwidth --- where many physical effects are most well measured --- could be affected. While the mass parameters are typically well measured from lower frequencies, the effects of spin and tidal parameters on the waveform phasing begin to become measurable in this region. Relative to the length-in-band of GW170817 ($\sim$ 100 s from 25 Hz to merger), the glitch duration is small (the overflow itself is less than 5~ms), as is the excision duration ($\sim$ 0.5~s). Ideally, handling this excision would require a modification of the signal template to similarly excise the portion of the signal which is removed with the data. While not conceptually complex, adding this capability to the existing low latency GW searches would increase computational costs.

\begin{figure*}
\begin{center}
\includegraphics[width=0.49\textwidth,clip=true]{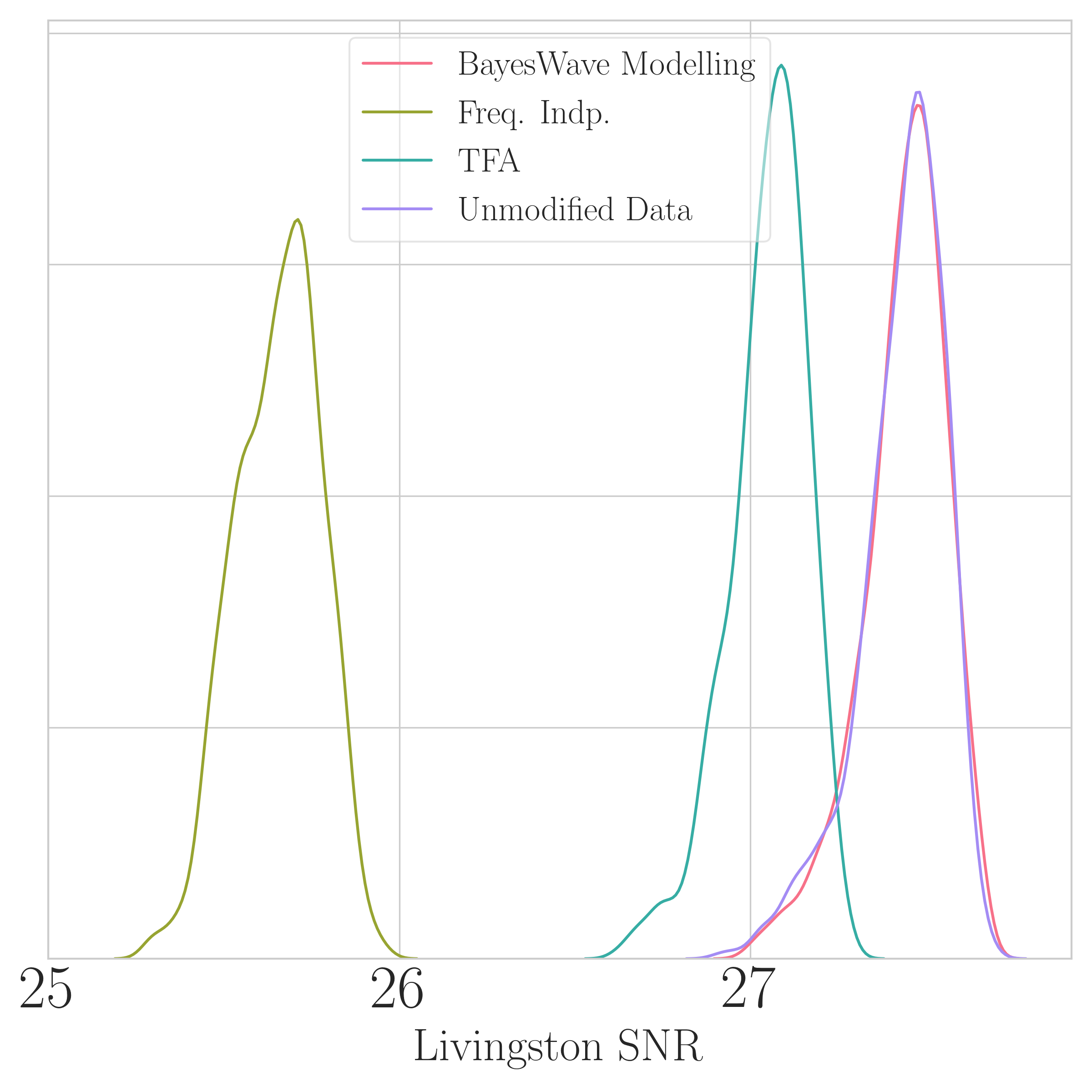}
\includegraphics[width=0.49\textwidth,clip=true]{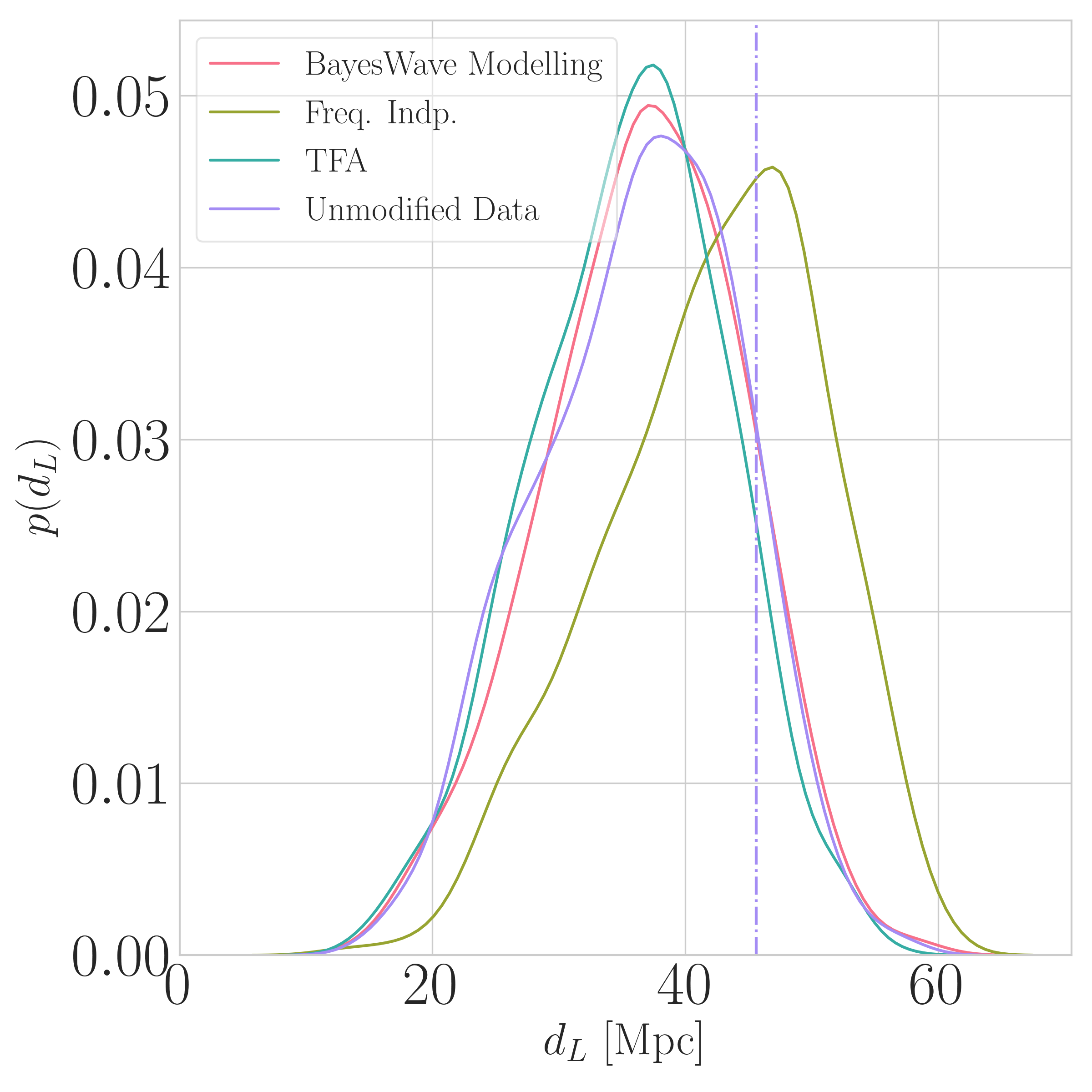}
\\
\includegraphics[width=0.49\textwidth,clip=true]{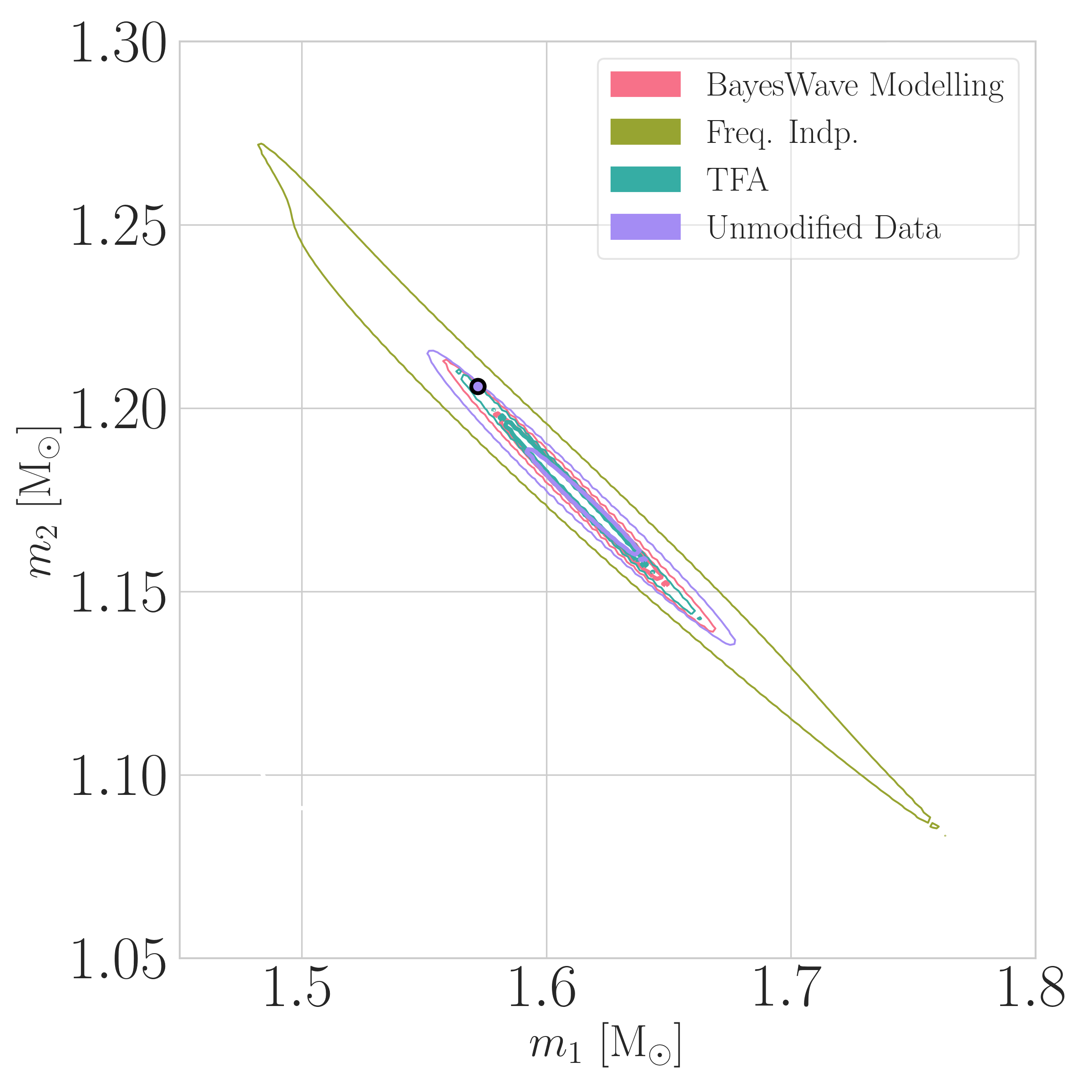}
\includegraphics[width=0.49\textwidth,clip=true]{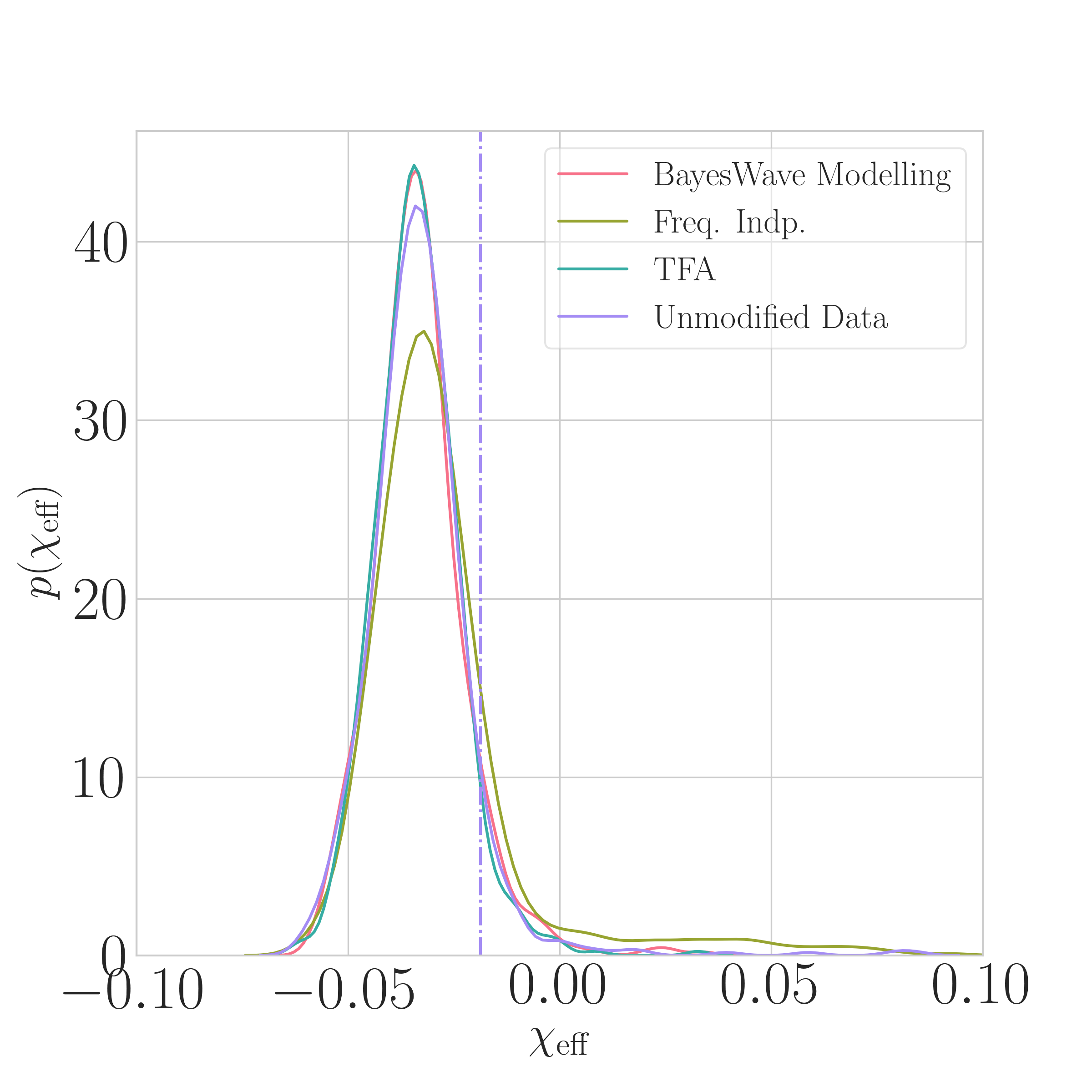}
\caption{\label{fig:phase_one_posteriors}
Comparison of mitigation methods. We show posterior distributions for the SNR in LIGO-Livingston (top left), the signal distance (top right), the component masses (bottom left), and the effective spin (bottom right) for a simulated signal injected into data adjacent in time to GW170817, but without a strong glitch near the end time. The posteriors are computed after applying the frequency-independent excision method of Sec.~\ref{sec:tukey_sub} (yellow, ``Freq.~Indp."), the frequency-dependent excision method of Sec.~\ref{sec:tf_sub} (purple, ``TFA"), after modeling and subtracting the glitch as described in Sec.~\ref{sec:bw_sub} (pink, ``BayesWave Modelling"), and when analysing the signal without any additional modification (green, ``No Gating"). The two excision methods lead to small biases in the estimation of the signal parameter. Moreover, the corresponding uncertainties are increased due to signal power being removed.}
\end{center}
\end{figure*}

For this example, no significant glitching is present near the merger time of the injection. The methods in Section \ref{sec:tukey_sub} and \ref{sec:tf_sub} each remove a fixed time-frequency volume from the data, and so the impact on the SNR of the signal should be the same with or without a glitch present. For the {\tt BayesWave} method in Section IIC, the absence of a glitch means that only a small amount of power is expected to be removed in this case.
The comparison between the different glitch-mitigation methods is displayed in Fig.~\ref{fig:phase_one_posteriors}. We present posterior distributions for various parameters of a signal injected, and the various mitigation techniques presented in Sec.~\ref{sec:glitch_removal} applied.
The SNR computed from data $s$ and template $h(\vec{\lambda}')$ is proportional to $\langle\tilde{s}|\tilde{h}(\vec{\lambda}')\rangle$, and as such, it is clear that removing some part of $s$ will incur a systematic reduction in the SNR distribution over the parameter set. 
This effect is apparent as the SNR distribution in the LIGO-Livingston data is reduced by about two units ($\sim7$\%) in the case of frequency-independent data excision and less than one unit ($\sim2$\%) for the frequency-dependent subtraction method (top left panel). The distance distribution is similarly modified as it is inversely proportional to the SNR, and the event appears to be slightly further away than in reality (top right panel). The distance / inclination degeneracy then leads to a marginal concurrent effect on the joint posterior for these two parameters.
Conversely, the SNR and distance distributions after the application of the {\tt BayesWave} glitch subtraction are very similar to the ones obtained when the signal is injected but otherwise unmodified.
This is the expected behavior, since there is no glitch to remove.

Also in Fig.~\ref{fig:phase_one_posteriors}, we show the effect of excision methods on the recovered detector-frame component masses (bottom left panel), the effective spin parameter\footnote{The effective spin parameter is the mass-weighted sum of the binary components' spins along the orbital angular momentum~\cite{Racine:2008qv} and one of the best measured spin combinations with GW data.} $\chi_{\textrm{eff}}$  (bottom right panel). 
The two dimensional marginalized posterior for the component masses shows that mass posteriors derived with different mitigation methods are broadly consistent with each other and give similar credible intervals along the small axis (chirp mass) of the posterior ellipse. However, the data excision methods lead to mass posteriors that encompass a broadened set of component masses and mass ratios. 
The widened posterior is indicative not only of the loss of SNR due to the flat portion of the window which entirely excises the data. The tapered portions of the window also induce spurious features in the Fourier domain which can serve to change the likelihood itself, altering the matching phase and amplitude of the data to a putative waveform.

This difference in the mass ratio estimate could affect the measurement of the spin parameters, due to the well-known spin-mass correlation~\cite{1994PhRvD..49.2658C}. Indeed, the posteriors for the effective spin parameter, shown in the bottom right plot of Fig.~\ref{fig:phase_one_posteriors}, exhibit a minor tail in the posteriors derived from excised data, reflecting the larger mass ratio uncertainty exhibited by the frequency independent and TFA excision data.

Overall, we find that both data excision methods recover a smaller SNR and, in the case of the frequency-independent method, can introduce small parameter biases, which are largely absent when the glitch is modeled and removed. These simple excision methods are an appealing option for quick glitch mitigation, due to their applicability to a large range of glitch morphologies, especially when rapid sky-localization is necessary.
However, for follow-up studies the glitch-fitting method seems preferable, since it never introduce biases, and can deal with cases where the glitch is of comparable length to the signal. In the next section we examine the performances of the glitch-fitting method using simulated signals.

\subsection{Efficacy and veracity of transient removal}
\label{sec:glitch_pe}

In the previous subsection we demonstrated that simple data excision techniques can lead to rapid source characterization with a tolerable degree of parameter biases. However, the recovered signal SNR is decreased, which implies that useful information about the signal has been lost. In this section we study in more detail the technique presented in Sec.~\ref{sec:bw_sub}, which attempts to fit the glitch and remove it while preserving the underlying signal and the Gaussian detector noise.  

\begin{figure*}
\begin{center}
\includegraphics[width=0.24\textwidth,clip=true]{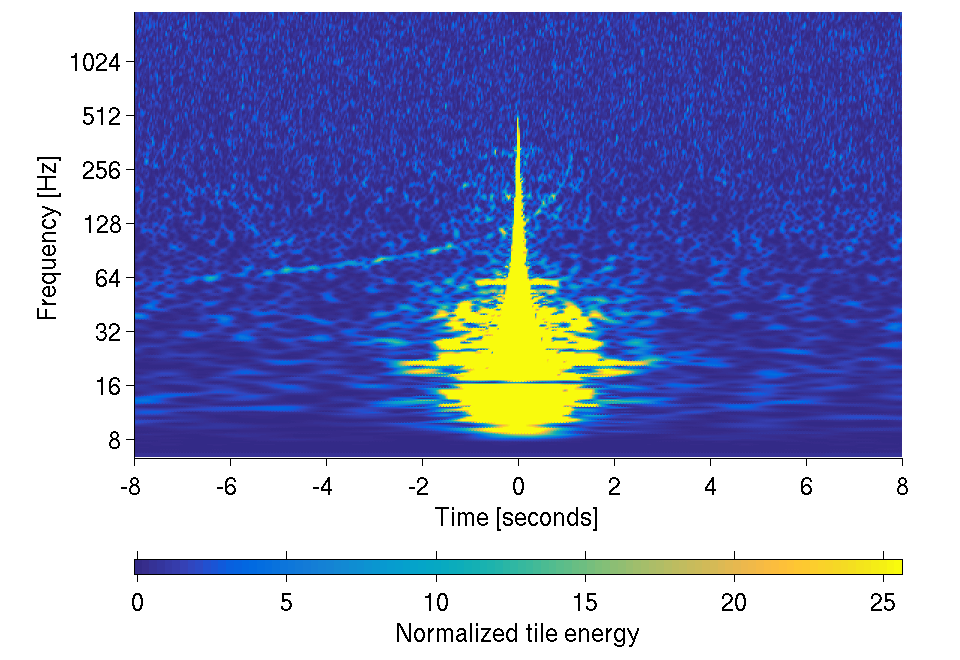}
\includegraphics[width=0.24\textwidth,clip=true]{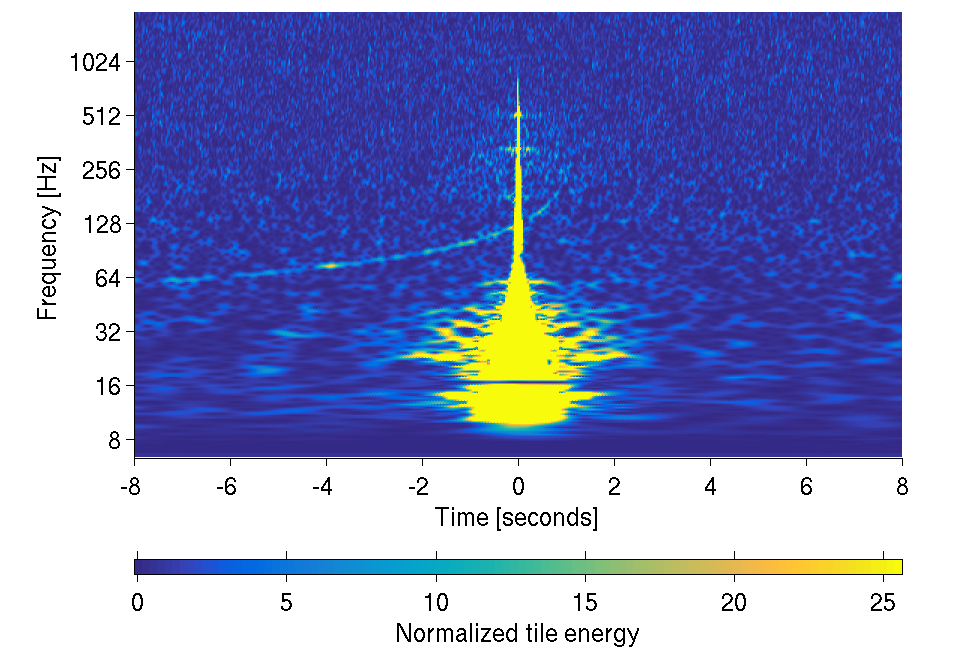}
\includegraphics[width=0.24\textwidth,clip=true]{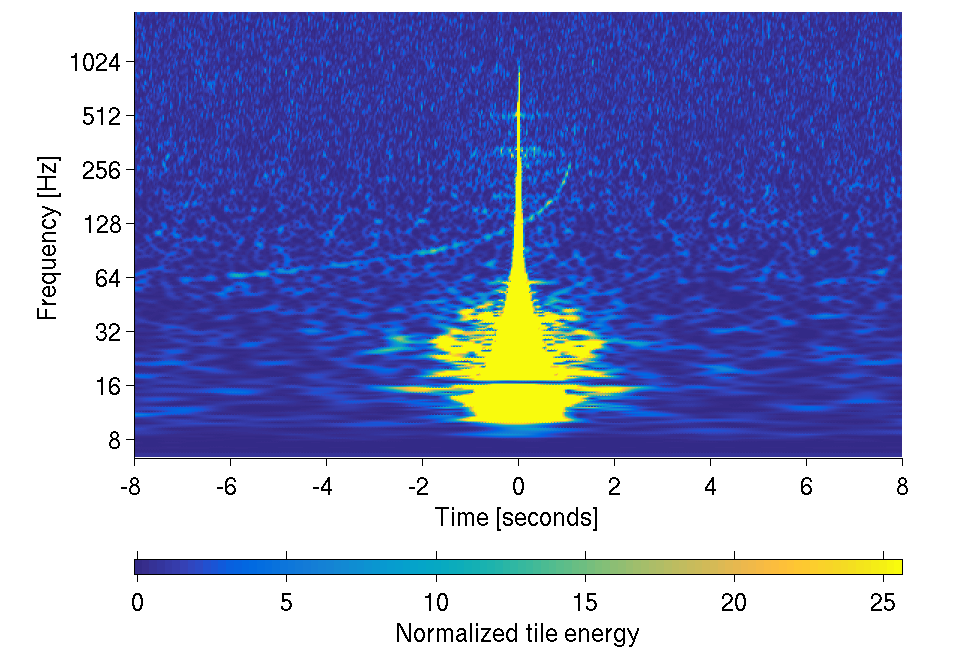}
\includegraphics[width=0.24\textwidth,clip=true]{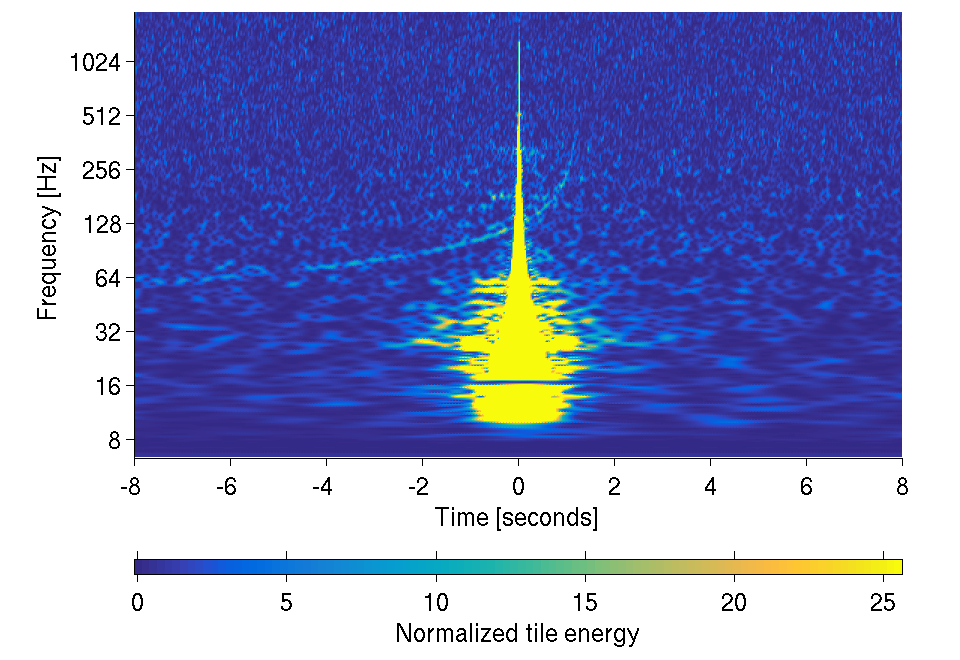} \\

\includegraphics[width=0.24\textwidth,clip=true]{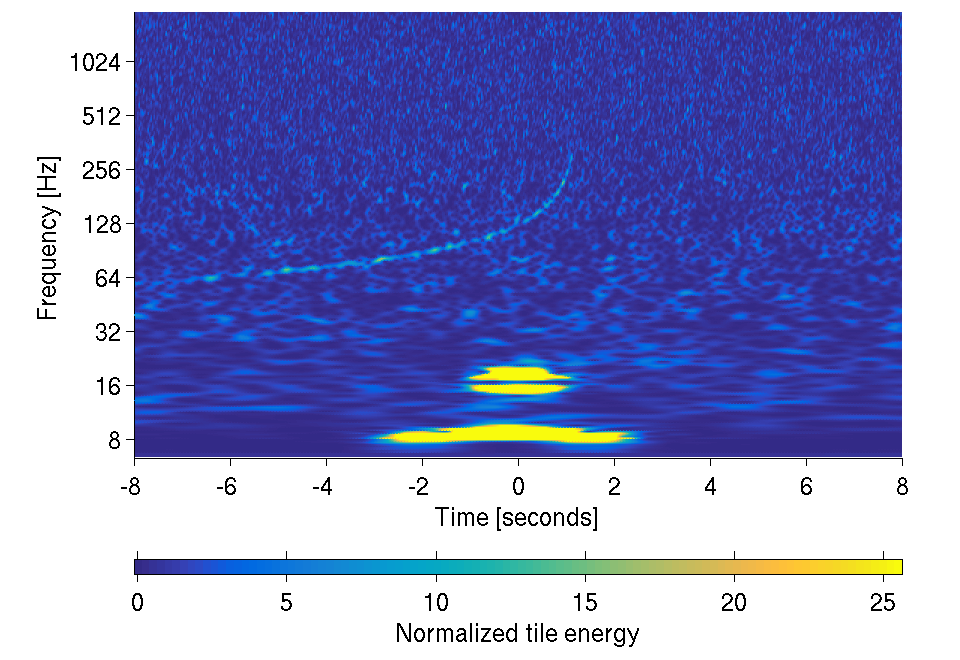}
\includegraphics[width=0.24\textwidth,clip=true]{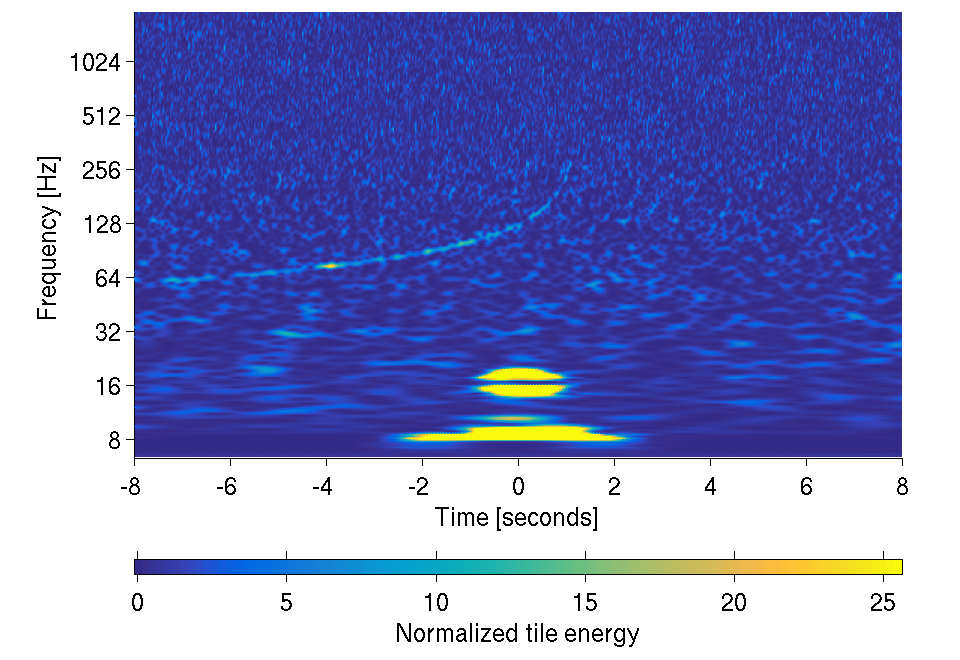}
\includegraphics[width=0.24\textwidth,clip=true]{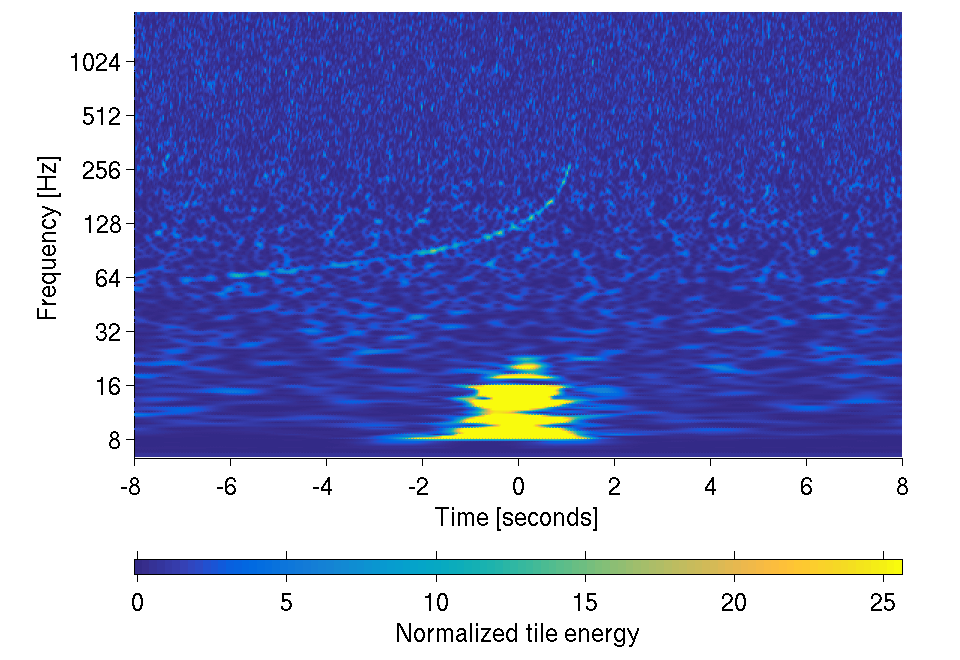}
\includegraphics[width=0.24\textwidth,clip=true]{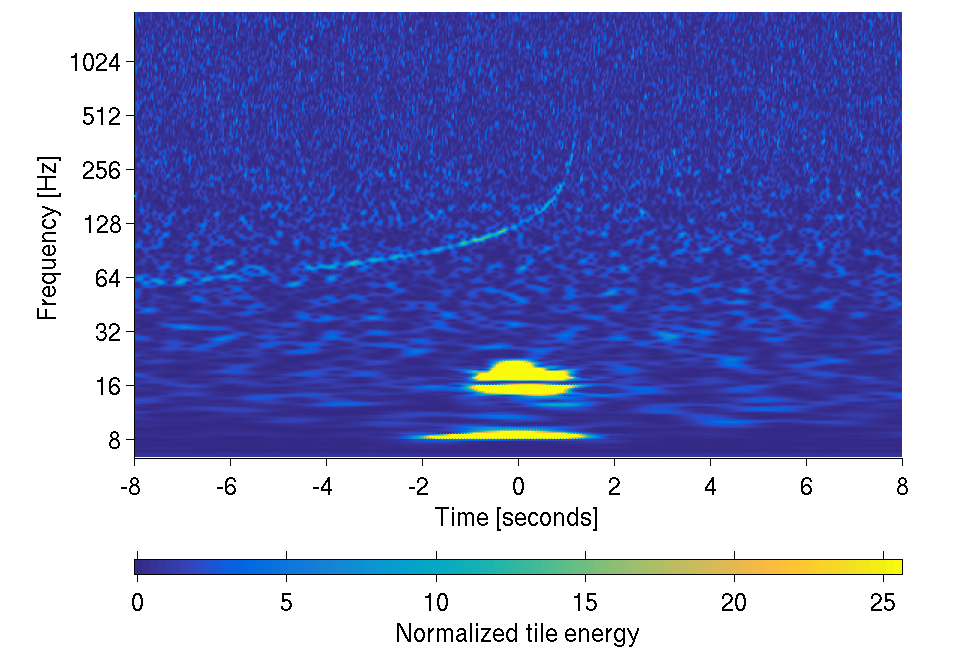}
\caption{\label{fig:glitch_TF}
Normalized signal amplitude (color axis) in the Livingston instrument for four synthetic events added to real interferometric data, as examined in a time-frequency representation. The top row is before the glitch-mitigation is applied, and the bottom row is after we have modeled and removed the glitch with {\tt BayesWave}. The simulated chirping signal is clearly visible in the background of all. The glitch extends in frequency below the lowest frequency used in either the glitch fitting and removal or the parameter estimation procedures, and as such was not completely removed. However, since those frequencies are excluded from the analyses, the result is unaffected by the remaining glitch power.} \end{center}
\end{figure*}

We produce GW170817-like signals and add them to the data of the LIGO-Livingston detector around four overflow glitches. To produce the signals we use the {\tt IMRPhenomPv2}~\cite{Hannam:2013oca,Schmidt:2014iyl,Husa:2015iqa,Khan:2015jqa} and the {\tt TaylorF2}~\cite{2002PhRvD..66b7502D} (including tidal corrections~\cite{2011PhRvD..83h4051V}) waveform models\footnote{When our study was started, no readily available waveform family which allows for both spin-precessional and finite-size tidal effects existed, though such a model has recently become available~\cite{Dietrich:2018uni}.} and we place the signals with end times 1.2 seconds after the time assigned to the Livingston overflow glitch we are studying. Table~\ref{table:injections} provides details for these simulated signals. We follow the same procedure from Ref.~\cite{TheLIGOScientific:2017qsa} in applying {\tt BayesWave} to model and subtract the glitch from the Livingston data following the method described in Sec.~\ref{sec:bw_sub}. Figure~\ref{fig:glitch_TF} shows the time-frequency representation of each glitch together with the simulated signal before and after glitch subtraction. The simulated signal is clearly visible as a characteristic `chirp' in both cases.

\begin{figure*}
\begin{center}
\includegraphics[width=0.3\textwidth,clip=true]{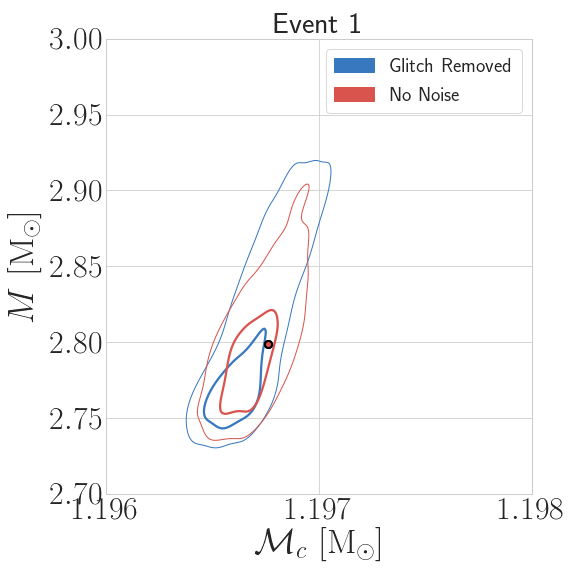}
\includegraphics[width=0.3\textwidth,clip=true]{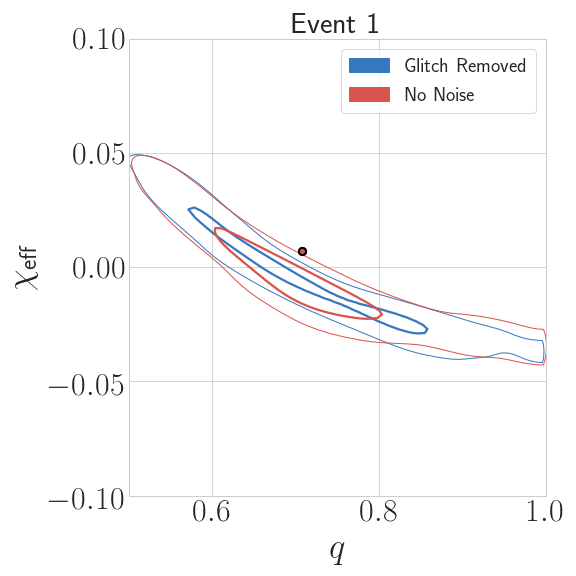}
\includegraphics[width=0.3\textwidth,clip=true]{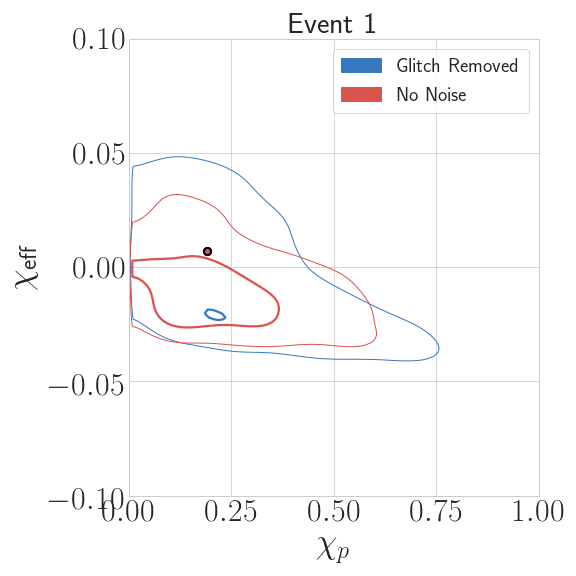}
\\
\includegraphics[width=0.3\textwidth,clip=true]{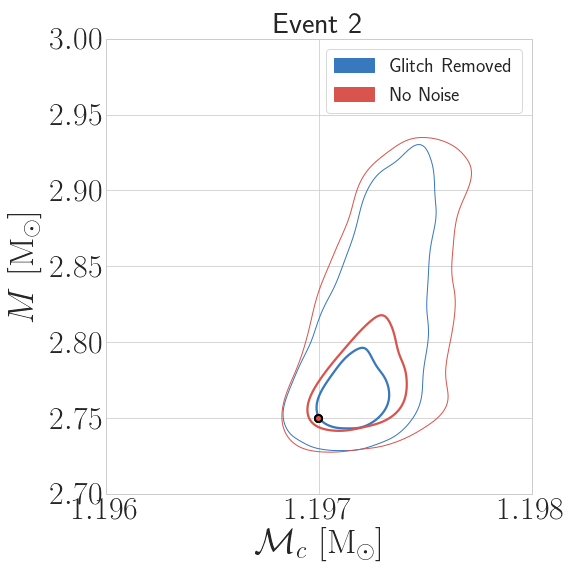}
\includegraphics[width=0.3\textwidth,clip=true]{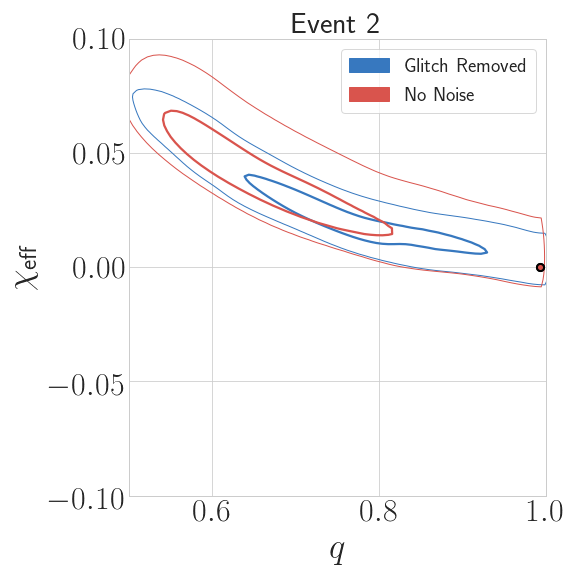}
\includegraphics[width=0.3\textwidth,clip=true]{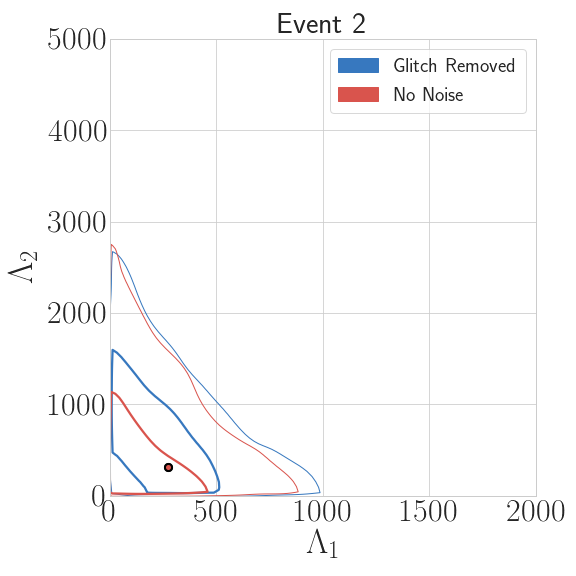}
\\
\includegraphics[width=0.3\textwidth,clip=true]{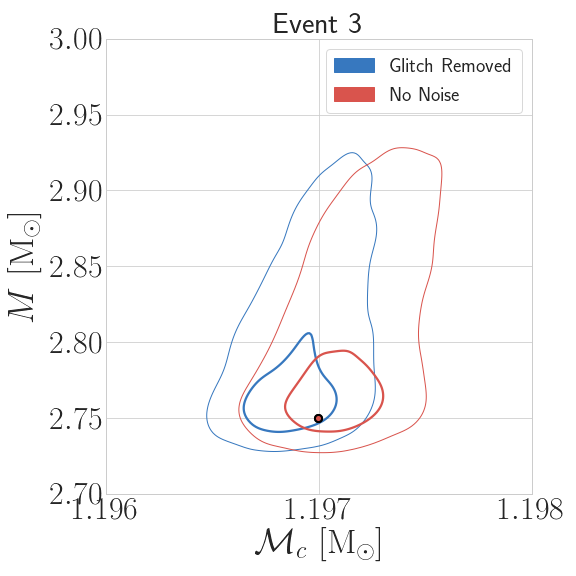}
\includegraphics[width=0.3\textwidth,clip=true]{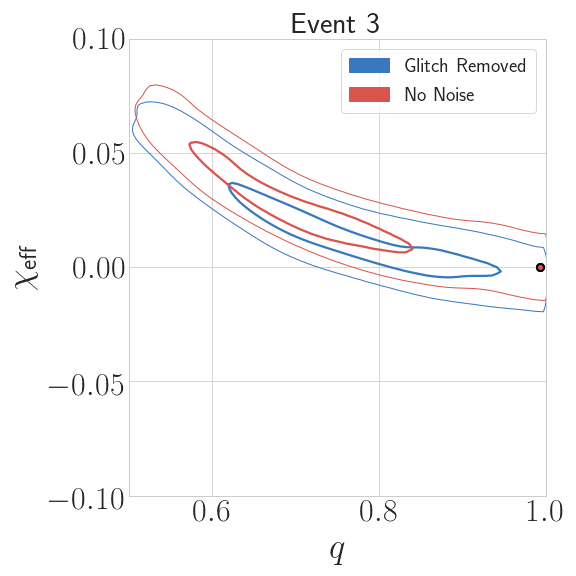}
\includegraphics[width=0.3\textwidth,clip=true]{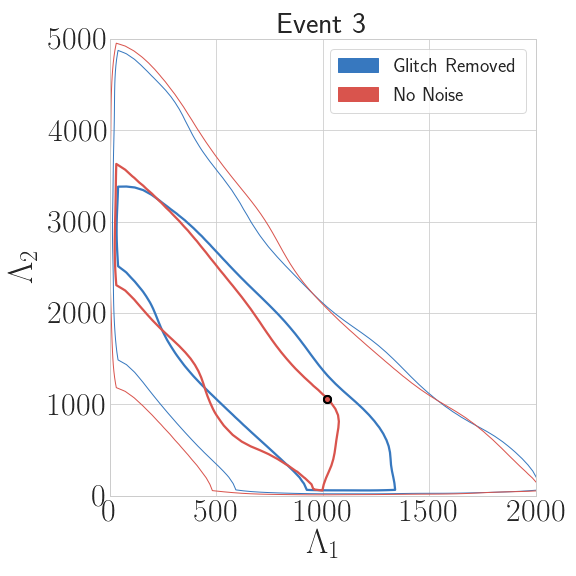}
\\
\includegraphics[width=0.3\textwidth,clip=true]{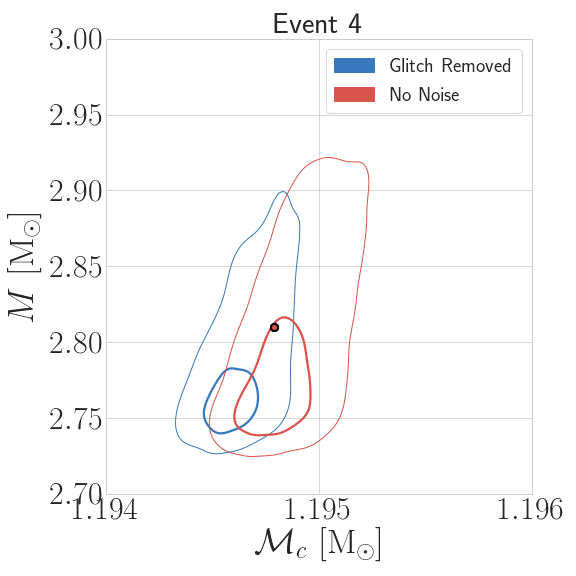}
\includegraphics[width=0.3\textwidth,clip=true]{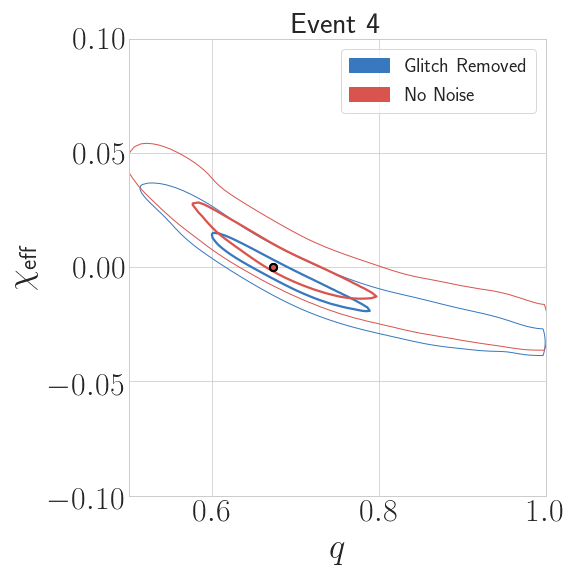}
\includegraphics[width=0.3\textwidth,clip=true]{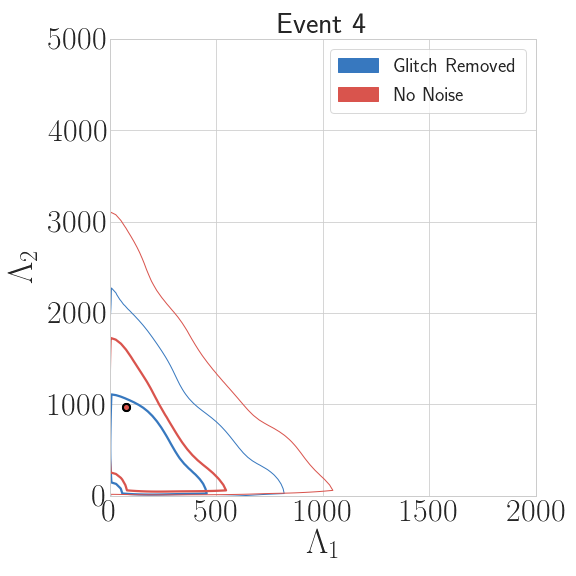}

\caption{\label{fig:pos_densities}
Posterior densities for the masses, spins, and tidal deformabilities for four overflow glitches (top to bottom). The left column shows the posteriors in total mass and chirp mass (in the detector frame); the middle column shows mass ratio versus $\chi_{\textrm{eff}}$; the last column shows $\chi_{\textrm{eff}}$ versus $\chi_{\textrm{p}}$ (first event, top row) or component tidal parameter posteriors $\Lambda_1$ and $\Lambda_2$ (remaining three events). Thick (thin) lines show the 50\% (90\%) credible regions. The true values of the parameters are represented by the black edged marker. In all cases the parameter estimates after removing the overflow glitch are consistent with estimates from injections in zero noise.}
\end{center}
\end{figure*}

We then perform a coherent Bayesian analysis to estimate the posterior probability distribution of the physical parameters of the simulated signals using the parameter estimation code {\tt LALInference}~\cite{Veitch:2014wba}. We obtain results by analyzing three instances of the synthetic event: (i) real data including the glitch, (ii) real data where the glitch has been subtracted, and (iii) simulated data with a zero-noise realization~\footnote{A zero-noise realization run assumes that the noise is exactly zero at all frequencies, while the power spectral density stays finite. This is a tool often used in the gravitational-wave literature, for example~\cite{Littenberg:2012uj,2014ApJ...784..119R,PhysRevLett.121.021303}, to assess parameter estimation in an controlled environment. Statistically, results obtained with a zero-noise realization are equivalent to the average results that would be obtained with a large sample of runs on Gaussian noise~\cite{Nissanke:2009kt,Vallisneri:2011ts}.}. In all cases, the power spectral density used to compute the likelihood in Eq.~\eqref{eqn:likelihood} is fit to the data surrounding and including the event and glitch time~\cite{Littenberg:2014oda} to ensure an even comparison between different runs. Of particular astrophysical interest are the physical aspects of the binary measurable through the gravitational-wave signal: its component masses $(m_1, m_2)$ via total mass and mass ratio, mass-weighted spins $(\chi_{\text{eff}})$, and tidal deformabilities $(\Lambda_1, \Lambda_2)$. In order to check the efficacy of recovery for either, we used the \texttt{IMRPhenomPv2} family to synthesize and insert three examples of a precessing signal, and the \texttt{TaylorF2} to insert three tidally influenced, but spin-aligned, waveforms into the data. In Fig.~\ref{fig:pos_densities} we present a selection of two-dimensional posteriors over combinations of the parameters for a precessing signal simulated with \texttt{IMRPhenomPv2} (top row) and three signals simulated with \texttt{TaylorF2} (second to fourth row). In all cases, we use the same waveform family to simulate the signal added into the data and as template in the parameter estimation algorithm.

In all four events shown in Fig.~\ref{fig:pos_densities}, we verify the recovery of the detector-frame mass parameters, since they dominate the phase evolution of the waveform. The left column of Fig.~\ref{fig:pos_densities} displays the total mass -- chirp mass posterior for all four simulated signals, with the zero-noise-realization and glitch-removed cases shown. The glitch-present recovery is often badly biased, enough so that we do not include it in Fig.~\ref{fig:pos_densities} and instead refer to Table \ref{table:injections} to indicate their credible regions. The posteriors for the glitch removed and zero-noise cases are qualitatively very similar. From Fig.~\ref{fig:pos_densities} it is seen that the mass recovery is consistent with the injected value in all cases, and the zero-noise and glitch-removed recovery encompasses similar values. This validates the premise that the glitch removal method does not bias the lower PN order parameters which influence the waveform.

\begin{table*}
\begin{centering}
\begin{tabular}{cccccc}
\hline
\noalign{\smallskip}
 & & & \multicolumn{3}{c}{Recovered Values} \\
Glitch GPS time & & True Value & Zero-Noise & No Mitigation & With Mitigation \\
\hline
\hline
\noalign{\smallskip}
Event 1 $\left(1186164816\right)$  & \specialcell{$m_1~[\textrm{M}_{\odot}]$ \\ $m_2~[\textrm{M}_{\odot}]$ \\ $q$ \\ $\chi_{\text{eff}}$ \\ $\Lambda_1$ \\ $\Lambda_2$} & \specialcell{1.64 \\ 1.16 \\ 0.71 \\ 0.007 \\ 0 \\ 0} & \specialcell{${1.64}^{+0.24}_{-0.23}$ \\ ${1.16}^{+0.18}_{-0.13}$ \\ ${0.70}^{+0.24}_{-0.16}$ \\ ${-0.01}^{+0.04}_{-0.02}$ \\  -  \\  - } & \specialcell{${24.44}^{+2.62}_{-0.92}$ \\ ${2.79}^{+0.20}_{-0.92}$ \\ ${0.11}^{+0.02}_{-0.03}$ \\ ${0.38}^{+0.02}_{-0.69}$ \\  -  \\  - } & \specialcell{${1.64}^{+0.25}_{-0.23}$ \\ ${1.16}^{+0.17}_{-0.14}$ \\ ${0.70}^{+0.23}_{-0.17}$ \\ ${-0.01}^{+0.05}_{-0.03}$ \\  -  \\  - } \\ \hline
Event 2 $\left(1186691156\right)$ & \specialcell{$m_1~[\textrm{M}_{\odot}]$ \\ $m_2~[\textrm{M}_{\odot}]$ \\ $q$ \\ $\chi_{\text{eff}}$ \\ $\Lambda_1$ \\ $\Lambda_2$} & \specialcell{1.38 \\ 1.37 \\ 0.99 \\ 0 \\ 275 \\ 309} & \specialcell{${1.64}^{+0.26}_{-0.24}$ \\ ${1.16}^{+0.19}_{-0.14}$ \\ ${0.71}^{+0.26}_{-0.17}$ \\ ${0.04}^{+0.05}_{-0.03}$ \\ ${193}^{+529}_{-180}$ \\ ${503}^{+1516}_{-467}$} & \specialcell{${61.19}^{+0.17}_{-0.96}$ \\ ${8.19}^{+0.07}_{-0.13}$ \\ ${0.13}^{+0.00}_{-0.00}$ \\ ${0.88}^{+0.00}_{-0.01}$ \\ ${1429}^{+45}_{-14}$ \\ ${2621}^{+2051}_{-2222}$} & \specialcell{${1.60}^{+0.29}_{-0.20}$ \\ ${1.19}^{+0.16}_{-0.17}$ \\ ${0.75}^{+0.22}_{-0.20}$ \\ ${0.02}^{+0.05}_{-0.02}$ \\ ${248}^{+568}_{-227}$ \\ ${668}^{+1395}_{-599}$} \\ \hline
Event 3 $\left(1186885739\right)$ & \specialcell{$m_1~[\textrm{M}_{\odot}]$ \\ $m_2~[\textrm{M}_{\odot}]$ \\ $q$ \\ $\chi_{\text{eff}}$ \\ $\Lambda_1$ \\ $\Lambda_2$} & \specialcell{1.38 \\ 1.37 \\ 0.99 \\ 0 \\ 1018 \\ 1062} & \specialcell{${1.60}^{+0.28}_{-0.21}$ \\ ${1.18}^{+0.17}_{-0.16}$ \\ ${0.74}^{+0.23}_{-0.19}$ \\ ${0.02}^{+0.05}_{-0.03}$ \\ ${588}^{+1084}_{-525}$ \\ ${1566}^{+2461}_{-1411}$} & \specialcell{${40.18}^{+0.44}_{-0.14}$ \\ ${6.17}^{+0.04}_{-0.07}$ \\ ${0.15}^{+0.001}_{-0.001}$ \\ ${0.33}^{+0.02}_{-0.02}$ \\ ${4958}^{+40}_{-144}$ \\ ${5.00}^{+1.01}_{-1.40}$} & \specialcell{${1.59}^{+0.28}_{-0.19}$ \\ ${1.20}^{+0.16}_{-0.17}$ \\ ${0.75}^{+0.21}_{-0.20}$ \\ ${0.01}^{+0.05}_{-0.02}$ \\ ${645}^{+1070}_{-574}$ \\ ${1545}^{+2420}_{-1382}$} \\ \hline
Event 4 $\left(1186300855\right)$ & \specialcell{$m_1~[\textrm{M}_{\odot}]$ \\ $m_2~[\textrm{M}_{\odot}]$ \\ $q$ \\ $\chi_{\text{eff}}$ \\ $\Lambda_1$ \\ $\Lambda_2$} & \specialcell{1.68 \\ 1.13 \\ 0.67 \\ 0 \\ 77 \\ 973} & \specialcell{${1.63}^{+0.25}_{-0.23}$ \\ ${1.16}^{+0.18}_{-0.14}$ \\ ${0.71}^{+0.25}_{-0.17}$ \\ ${0}^{+0.04}_{-0.03}$ \\ ${240}^{+605}_{-220}$ \\ ${728}^{+1686}_{-668}$} & \specialcell{${38.71}^{+0.08}_{-0.05}$ \\ ${5.80}^{+0.02}_{-0.02}$ \\ ${0.15}^{+0.001}_{-0.001}$ \\ ${0.33}^{+0.01}_{-0.01}$ \\ ${4991}^{+8}_{-27}$ \\ ${0.14}^{+0.02}_{-0.03}$} & \specialcell{${1.62}^{+0.26}_{-0.22}$ \\ ${1.17}^{+0.17}_{-0.15}$ \\ ${0.72}^{+0.23}_{-0.18}$ \\ ${-0.01}^{+0.04}_{-0.02}$ \\ ${211}^{+576}_{-193}$ \\ ${591}^{+1502}_{-539}$} \\ \hline
\noalign{\smallskip}
\end{tabular}
\end{centering}
\caption{Properties of the four simulated signals analyzed. The first column gives the GPS time of the overflow glitch around which we make the BNS injections; the second column list the binary parameters we study; the third column gives they injected (true values). The remaining three columns give the median recovered value for each parameter, as well as the 90\% credible intervals in the case of a zero-noise-realization injection (fourth column), a parameter estimation analysis on data that include the glitch (fifth column) and on data obtained after glitch mitigation with {\tt BayesWave} (sixth column).}
\label{table:injections}
\end{table*}

The second column of Fig.~\ref{fig:pos_densities} shows the two-dimensional posteriors for the mass ratio and the effective spin. We again find no bias in the parameter estimates due to the glitch removal. The posteriors for the glitch removal and the zero-noise analyses are minimally shifted with respect to each other, which is consistent with the expected effect of noise realization on signal recovery and not evidence for a bias. Indeed, the specific noise realization of the data induces an additional shift on the posterior estimates of the order of the posterior variance. In addition, the top-right panel shows the posterior for the effective spin $\chi_{\textrm{eff}}$ and the spin parameter $\chi_{\textrm{p}}$~\cite{Schmidt:2014iyl}. The latter parameter is an estimate of spin-precession in the waveform. As expected we again see no biases due to glitch removal.

Finally, the tidal parameter estimation also seems  unaffected by the glitch removal procedure. The two dimensional marginalized posteriors for the component tidal parameters are presented in the right column, second through fourth row of Fig.~\ref{fig:pos_densities}. The posteriors obtained with our two analyses are both consistent with each other and capture the known value well within their credible intervals. Moreover, the recovered posterior structure is very similar to the actual posterior measured for GW170817 in~\cite{TheLIGOScientific:2017qsa}, exhibiting similar boundaries and degeneracies.

Credible intervals for the parameters, as well as their injected values, are quoted in Table \ref{table:injections} for all three analyses (with glitch mitigation, without glitch mitigation, and with a zero-noise realization). The $90\%$ credible intervals computed from the glitch-mitigated data are consistent with the ones from the zero-noise analysis. This is in stark contrast to the parameter estimates computed if the glitch is included in the data analyzed. The values recovered are well outside of the posterior in the glitch-free examples, and nowhere near the known values. In all four cases, the mass ratio is pushed to extremely high values. The same extreme displacement occurs for the tidal parameters, producing tidal deformability values in the thousands, strongly peaked for $\Lambda_1$ and nearly unmeasurable for $\Lambda_2$. The uncertainties from the posteriors are wider than their glitch-free counterparts. Despite this, the credible intervals miss the known value by a wide margin, several times their own width. This implies that without mitigation, not only is the most basic parameter estimation biased, but simply incorrect. 

\section{Conclusions}
\label{sec:conl}

The focus of this work is to examine potential biases induced by various noise transient subtraction methods. There is a trade off between speed and fidelity in terms of their efficacy. Simple but fast methods which ignore the morphology of the noise transient or the data spectra are likely to introduce noticeable and predictable biases in signal amplitude recovery. Also worth noting is that the bias and widening of uncertainty induced by these methods may be more severe for weaker (lower SNR) signals --- GW170817 was the highest SNR GW signal detected to date. Amelioration of these biases by introducing the same gating procedure to the template waveform remains impractical. In a low latency environment where sky maps are required without waiting for a full pass of a glitch extraction method like {\tt BayesWave}, the time-frequency data excision in Section \ref{sec:tf_sub} seems best suited and of appropriate latency. While parameter biases are reduced for more careful treatments of data excision, they are not a substitute for full glitch subtraction --- particularly when high data fidelity is required to retrieve an accurate estimation of physical parameters. Indeed we find that the data produced after glitch-mitigation by {\tt BayesWave} are consistent with our zero-noise study to within shifts that can be attributed to the presence of noise.

The consistent recovery of the spin and tidal parameters are important tests: the frequencies affected are not only in the most sensitive bandwidth of the GW interferometers, but also where subdominant effects on the phase of the waveform, such as those from spin and tidal deformabilities, become measurable. If the glitch removal procedure damaged the structure or coherence of the signal, then it would be expected that those parameters would exhibit significant bias. A study examining the effect of glitches on source property estimation by time-shifting glitches along the time-frequency track of signals is in preparation~\cite{McIver2018}. Another study which examines the biases in physical parameter recovery for binary black hole signals where the merger time overlaps a glitch can be found in~\cite{Powell:2018csz}.

The overflow glitch studied here is, in some sense, a best case scenario for glitch subtraction with {\tt BayesWave}, because the signal and glitch have sufficiently different morphology. In particular, while the signal and glitch overlap in time-frequency (T-F) space, they are fit by wavelets having very different quality factor Q (which governs the number of cycles in the wavelet)--high Q for the BNS and low Q for the glitch. Generally, when glitches and signals do not overlap in T-F-Q space, a straightforward application of BayesWave will provide clean residuals for the PE analysis. Fortunately, that was the scenario presented by GW170817. Higher mass GW signals, such as BBH mergers, use lower Q wavelets and would therefore be more difficult to separate from a similar glitch with the current analysis tools. To mitigate against such a scenario in the future, we are developing glitch subtraction algorithms which include information about the network coherence of the signal, and lack thereof for any glitches.

Since multi-messenger astronomy involving NSs in compact binaries is expected to be highly featured in the upcoming Advanced LIGO/Virgo/KAGRA observing runs~\cite{2016LRR....19....1A}, confidence in the measured properties of GW signals is paramount. In the presence of short-duration noise transients near a merger signal, gating and data subtraction procedures are acceptable stopgap measures to obtain data products needed for rapid identification of electromagnetic counterparts. However, to obtain unbiased results without inflating uncertainties, we find that data-tailored glitch-removal techniques such as { \tt BayesWave} are required.

\acknowledgments
The authors are grateful for useful conversations with Duncan Brown and Stanislav Babak in the construction of the manuscript. CP is supported by the NSF grant PHY-1607709, and also acknowledges support by the Center for Interdisciplinary Exploration and Research in Astrophysics (CIERA). EAC thanks the LSSTC Data Science Fellowship Program; her time as a fellow has benefited this work. EAC also acknowledges support from the Illinois Space Grant Consortium Graduate Fellowship Program. NC acknowledges support from the NSF award PHY-1306702.
ME and SV acknowledge the support of the National Science Foundation and the LIGO Laboratory. LIGO was constructed by the California Institute of Technology and Massachusetts Institute of Technology with funding from the National Science Foundation and operates under cooperative agreement PHY-0757058.

The authors thank the LIGO Scientific Collaboration for access to the data and gratefully acknowledge the support of the United States National Science Foundation (NSF) for the construction and operation of the LIGO Laboratory and Advanced LIGO as well as the Science and Technology Facilities Council (STFC) of the United Kingdom, and the Max-Planck-Society (MPS) for support of the construction of Advanced LIGO. Additional support for Advanced LIGO was provided by the Australian Research Council.

\bibliography{Refs}

\begin{thebibliography}{52}
\expandafter\ifx\csname natexlab\endcsname\relax\def\natexlab#1{#1}\fi
\expandafter\ifx\csname bibnamefont\endcsname\relax
  \def\bibnamefont#1{#1}\fi
\expandafter\ifx\csname bibfnamefont\endcsname\relax
  \def\bibfnamefont#1{#1}\fi
\expandafter\ifx\csname citenamefont\endcsname\relax
  \def\citenamefont#1{#1}\fi
\expandafter\ifx\csname url\endcsname\relax
  \def\url#1{\texttt{#1}}\fi
\expandafter\ifx\csname urlprefix\endcsname\relax\def\urlprefix{URL }\fi
\providecommand{\bibinfo}[2]{#2}
\providecommand{\eprint}[2][]{\url{#2}}

\bibitem[{\citenamefont{Abbott
  et~al.}(2017{\natexlab{a}})}]{TheLIGOScientific:2017qsa}
\bibinfo{author}{\bibfnamefont{B.~P.} \bibnamefont{Abbott}}
  \bibnamefont{et~al.} (\bibinfo{collaboration}{{Virgo, LIGO Scientific}}),
  \bibinfo{journal}{Phys. Rev. Lett.} \textbf{\bibinfo{volume}{119}},
  \bibinfo{pages}{161101} (\bibinfo{year}{2017}{\natexlab{a}}),
  \eprint{1710.05832}.

\bibitem[{\citenamefont{Abbott et~al.}(2017{\natexlab{b}})}]{Monitor:2017mdv}
\bibinfo{author}{\bibfnamefont{B.~P.} \bibnamefont{Abbott}}
  \bibnamefont{et~al.} (\bibinfo{collaboration}{Virgo, Fermi-GBM, INTEGRAL,
  LIGO Scientific}), \bibinfo{journal}{Astrophys. J.}
  \textbf{\bibinfo{volume}{848}}, \bibinfo{pages}{L13}
  (\bibinfo{year}{2017}{\natexlab{b}}), \eprint{1710.05834}.

\bibitem[{\citenamefont{Abbott et~al.}(2017{\natexlab{c}})}]{GBM:2017lvd}
\bibinfo{author}{\bibfnamefont{B.~P.} \bibnamefont{Abbott}}
  \bibnamefont{et~al.} (\bibinfo{collaboration}{GROND, SALT Group, OzGrav, DFN,
  INTEGRAL, Virgo, Insight-Hxmt, MAXI Team, Fermi-LAT, J-GEM, RATIR, IceCube,
  CAASTRO, LWA, ePESSTO, GRAWITA, RIMAS, SKA South Africa/MeerKAT, H.E.S.S.,
  1M2H Team, IKI-GW Follow-up, Fermi GBM, Pi of Sky, DWF (Deeper Wider Faster
  Program), Dark Energy Survey, MASTER, AstroSat Cadmium Zinc Telluride Imager
  Team, Swift, Pierre Auger, ASKAP, VINROUGE, JAGWAR, Chandra Team at McGill
  University, TTU-NRAO, GROWTH, AGILE Team, MWA, ATCA, AST3, TOROS, Pan-STARRS,
  NuSTAR, ATLAS Telescopes, BOOTES, CaltechNRAO, LIGO Scientific, High Time
  Resolution Universe Survey, Nordic Optical Telescope, Las Cumbres Observatory
  Group, TZAC Consortium, LOFAR, IPN, DLT40, Texas Tech University, HAWC,
  ANTARES, KU, Dark Energy Camera GW-EM, CALET, Euro VLBI Team, ALMA}),
  \bibinfo{journal}{Astrophys. J.} \textbf{\bibinfo{volume}{848}},
  \bibinfo{pages}{L12} (\bibinfo{year}{2017}{\natexlab{c}}),
  \eprint{1710.05833}.

\bibitem[{\citenamefont{Aasi et~al.}(2015)}]{TheLIGOScientific:2014jea}
\bibinfo{author}{\bibfnamefont{J.}~\bibnamefont{Aasi}} \bibnamefont{et~al.}
  (\bibinfo{collaboration}{LIGO Scientific}), \bibinfo{journal}{Class. Quant.
  Grav.} \textbf{\bibinfo{volume}{32}}, \bibinfo{pages}{074001}
  (\bibinfo{year}{2015}), \eprint{1411.4547}.

\bibitem[{\citenamefont{Acernese et~al.}(2015)}]{TheVirgo:2014hva}
\bibinfo{author}{\bibfnamefont{F.}~\bibnamefont{Acernese}} \bibnamefont{et~al.}
  (\bibinfo{collaboration}{VIRGO}), \bibinfo{journal}{Class. Quant. Grav.}
  \textbf{\bibinfo{volume}{32}}, \bibinfo{pages}{024001}
  (\bibinfo{year}{2015}), \eprint{1408.3978}.

\bibitem[{\citenamefont{Messick et~al.}(2017)\citenamefont{Messick, Blackburn,
  Brady, Brockill, Cannon, Cariou, Caudill, Chamberlin, Creighton, Everett
  et~al.}}]{PhysRevD.95.042001}
\bibinfo{author}{\bibfnamefont{C.}~\bibnamefont{Messick}},
  \bibinfo{author}{\bibfnamefont{K.}~\bibnamefont{Blackburn}},
  \bibinfo{author}{\bibfnamefont{P.}~\bibnamefont{Brady}},
  \bibinfo{author}{\bibfnamefont{P.}~\bibnamefont{Brockill}},
  \bibinfo{author}{\bibfnamefont{K.}~\bibnamefont{Cannon}},
  \bibinfo{author}{\bibfnamefont{R.}~\bibnamefont{Cariou}},
  \bibinfo{author}{\bibfnamefont{S.}~\bibnamefont{Caudill}},
  \bibinfo{author}{\bibfnamefont{S.~J.} \bibnamefont{Chamberlin}},
  \bibinfo{author}{\bibfnamefont{J.~D.~E.} \bibnamefont{Creighton}},
  \bibinfo{author}{\bibfnamefont{R.}~\bibnamefont{Everett}},
  \bibnamefont{et~al.}, \bibinfo{journal}{Phys. Rev. D}
  \textbf{\bibinfo{volume}{95}}, \bibinfo{pages}{042001}
  (\bibinfo{year}{2017}),
  \urlprefix\url{https://link.aps.org/doi/10.1103/PhysRevD.95.042001}.

\bibitem[{\citenamefont{{Nitz} et~al.}(2018)\citenamefont{{Nitz}, {Dal Canton},
  {Davis}, and {Reyes}}}]{2018arXiv180511174N}
\bibinfo{author}{\bibfnamefont{A.~H.} \bibnamefont{{Nitz}}},
  \bibinfo{author}{\bibfnamefont{T.}~\bibnamefont{{Dal Canton}}},
  \bibinfo{author}{\bibfnamefont{D.}~\bibnamefont{{Davis}}}, \bibnamefont{and}
  \bibinfo{author}{\bibfnamefont{S.}~\bibnamefont{{Reyes}}},
  \bibinfo{journal}{ArXiv e-prints} \bibinfo{eid}{arXiv:1805.11174}
  (\bibinfo{year}{2018}), \eprint{1805.11174}.

\bibitem[{\citenamefont{Singer and Price}(2016)}]{PhysRevD.93.024013}
\bibinfo{author}{\bibfnamefont{L.~P.} \bibnamefont{Singer}} \bibnamefont{and}
  \bibinfo{author}{\bibfnamefont{L.~R.} \bibnamefont{Price}},
  \bibinfo{journal}{Phys. Rev. D} \textbf{\bibinfo{volume}{93}},
  \bibinfo{pages}{024013} (\bibinfo{year}{2016}),
  \urlprefix\url{https://link.aps.org/doi/10.1103/PhysRevD.93.024013}.

\bibitem[{\citenamefont{{Nuttall} et~al.}(2015)\citenamefont{{Nuttall},
  {Massinger}, {Areeda}, {Betzwieser}, {Dwyer}, {Effler}, {Fisher},
  {Fritschel}, {Kissel}, {Lundgren} et~al.}}]{2015CQGra..32x5005N}
\bibinfo{author}{\bibfnamefont{L.~K.} \bibnamefont{{Nuttall}}},
  \bibinfo{author}{\bibfnamefont{T.~J.} \bibnamefont{{Massinger}}},
  \bibinfo{author}{\bibfnamefont{J.}~\bibnamefont{{Areeda}}},
  \bibinfo{author}{\bibfnamefont{J.}~\bibnamefont{{Betzwieser}}},
  \bibinfo{author}{\bibfnamefont{S.}~\bibnamefont{{Dwyer}}},
  \bibinfo{author}{\bibfnamefont{A.}~\bibnamefont{{Effler}}},
  \bibinfo{author}{\bibfnamefont{R.~P.} \bibnamefont{{Fisher}}},
  \bibinfo{author}{\bibfnamefont{P.}~\bibnamefont{{Fritschel}}},
  \bibinfo{author}{\bibfnamefont{J.~S.} \bibnamefont{{Kissel}}},
  \bibinfo{author}{\bibfnamefont{A.~P.} \bibnamefont{{Lundgren}}},
  \bibnamefont{et~al.}, \bibinfo{journal}{Classical and Quantum Gravity}
  \textbf{\bibinfo{volume}{32}} (\bibinfo{year}{2015}).

\bibitem[{\citenamefont{{Abbott} et~al.}(2016)\citenamefont{{Abbott}, {Abbott},
  {Abbott}, {Abernathy}, {Acernese}, {Ackley}, {Adamo}, {Adams}, {Adams},
  {Addesso} et~al.}}]{2016CQGra..33m4001A}
\bibinfo{author}{\bibfnamefont{B.~P.} \bibnamefont{{Abbott}}},
  \bibinfo{author}{\bibfnamefont{R.}~\bibnamefont{{Abbott}}},
  \bibinfo{author}{\bibfnamefont{T.~D.} \bibnamefont{{Abbott}}},
  \bibinfo{author}{\bibfnamefont{M.~R.} \bibnamefont{{Abernathy}}},
  \bibinfo{author}{\bibfnamefont{F.}~\bibnamefont{{Acernese}}},
  \bibinfo{author}{\bibfnamefont{K.}~\bibnamefont{{Ackley}}},
  \bibinfo{author}{\bibfnamefont{M.}~\bibnamefont{{Adamo}}},
  \bibinfo{author}{\bibfnamefont{C.}~\bibnamefont{{Adams}}},
  \bibinfo{author}{\bibfnamefont{T.}~\bibnamefont{{Adams}}},
  \bibinfo{author}{\bibfnamefont{P.}~\bibnamefont{{Addesso}}},
  \bibnamefont{et~al.}, \bibinfo{journal}{Classical and Quantum Gravity}
  \textbf{\bibinfo{volume}{33}} (\bibinfo{year}{2016}).

\bibitem[{\citenamefont{{Abbott} et~al.}(2018)\citenamefont{{Abbott}, {Abbott},
  {Abbott}, {Abernathy}, {Acernese}, {Ackley}, {Adams}, {Adams}, {Addesso},
  {Adhikari} et~al.}}]{2018CQGra..35f5010A}
\bibinfo{author}{\bibfnamefont{B.~P.} \bibnamefont{{Abbott}}},
  \bibinfo{author}{\bibfnamefont{R.}~\bibnamefont{{Abbott}}},
  \bibinfo{author}{\bibfnamefont{T.~D.} \bibnamefont{{Abbott}}},
  \bibinfo{author}{\bibfnamefont{M.~R.} \bibnamefont{{Abernathy}}},
  \bibinfo{author}{\bibfnamefont{F.}~\bibnamefont{{Acernese}}},
  \bibinfo{author}{\bibfnamefont{K.}~\bibnamefont{{Ackley}}},
  \bibinfo{author}{\bibfnamefont{C.}~\bibnamefont{{Adams}}},
  \bibinfo{author}{\bibfnamefont{T.}~\bibnamefont{{Adams}}},
  \bibinfo{author}{\bibfnamefont{P.}~\bibnamefont{{Addesso}}},
  \bibinfo{author}{\bibfnamefont{R.~X.} \bibnamefont{{Adhikari}}},
  \bibnamefont{et~al.}, \bibinfo{journal}{Classical and Quantum Gravity}
  \textbf{\bibinfo{volume}{35}} (\bibinfo{year}{2018}).

\bibitem[{\citenamefont{{Usman} et~al.}(2016)\citenamefont{{Usman}, {Nitz},
  {Harry}, {Biwer}, {Brown}, {Cabero}, {Capano}, {Dal Canton}, {Dent},
  {Fairhurst} et~al.}}]{2016CQGra..33u5004U}
\bibinfo{author}{\bibfnamefont{S.~A.} \bibnamefont{{Usman}}},
  \bibinfo{author}{\bibfnamefont{A.~H.} \bibnamefont{{Nitz}}},
  \bibinfo{author}{\bibfnamefont{I.~W.} \bibnamefont{{Harry}}},
  \bibinfo{author}{\bibfnamefont{C.~M.} \bibnamefont{{Biwer}}},
  \bibinfo{author}{\bibfnamefont{D.~A.} \bibnamefont{{Brown}}},
  \bibinfo{author}{\bibfnamefont{M.}~\bibnamefont{{Cabero}}},
  \bibinfo{author}{\bibfnamefont{C.~D.} \bibnamefont{{Capano}}},
  \bibinfo{author}{\bibfnamefont{T.}~\bibnamefont{{Dal Canton}}},
  \bibinfo{author}{\bibfnamefont{T.}~\bibnamefont{{Dent}}},
  \bibinfo{author}{\bibfnamefont{S.}~\bibnamefont{{Fairhurst}}},
  \bibnamefont{et~al.}, \bibinfo{journal}{Classical and Quantum Gravity}
  \textbf{\bibinfo{volume}{33}} (\bibinfo{year}{2016}).

\bibitem[{\citenamefont{Nitz et~al.}(2017)\citenamefont{Nitz, Harry, Brown,
  Biwer, Willis, Canton et~al.}}]{pycbc-software}
\bibinfo{author}{\bibfnamefont{A.}~\bibnamefont{Nitz}},
  \bibinfo{author}{\bibfnamefont{I.}~\bibnamefont{Harry}},
  \bibinfo{author}{\bibfnamefont{D.}~\bibnamefont{Brown}},
  \bibinfo{author}{\bibfnamefont{C.~M.} \bibnamefont{Biwer}},
  \bibinfo{author}{\bibfnamefont{J.}~\bibnamefont{Willis}},
  \bibinfo{author}{\bibfnamefont{T.~D.} \bibnamefont{Canton}},
  \bibnamefont{et~al.}, \emph{\bibinfo{title}{ligo-cbc/pycbc: O2 production
  release 19}} (\bibinfo{year}{2017}),
  \urlprefix\url{https://zenodo.org/record/852372#.Wz-QcVMvwxM}.

\bibitem[{\citenamefont{{Metzger} and {Berger}}(2012)}]{2012ApJ...746...48M}
\bibinfo{author}{\bibfnamefont{B.~D.} \bibnamefont{{Metzger}}}
  \bibnamefont{and} \bibinfo{author}{\bibfnamefont{E.}~\bibnamefont{{Berger}}},
  \bibinfo{journal}{\apj} \textbf{\bibinfo{volume}{746}}
  (\bibinfo{year}{2012}).

\bibitem[{\citenamefont{{The {LIGO} Scientific Collaboration}
  et~al.}(2017)\citenamefont{{The {LIGO} Scientific Collaboration}, {The Virgo
  Collaboration} et~al.}}]{GCN21513}
\bibinfo{author}{\bibnamefont{{The {LIGO} Scientific Collaboration}}},
  \bibinfo{author}{\bibnamefont{{The Virgo Collaboration}}},
  \bibnamefont{et~al.}, \bibinfo{journal}{GCN}
  \textbf{\bibinfo{volume}{21513}}, \bibinfo{pages}{1} (\bibinfo{year}{2017}).

\bibitem[{\citenamefont{Coulter et~al.}(2017)}]{2017CoulterNat}
\bibinfo{author}{\bibfnamefont{D.~A.} \bibnamefont{Coulter}}
  \bibnamefont{et~al.}, \bibinfo{journal}{Science}
  \textbf{\bibinfo{volume}{358}}, \bibinfo{pages}{1556} (\bibinfo{year}{2017}),
  \eprint{1710.05452}.

\bibitem[{\citenamefont{Cornish and Littenberg}(2015)}]{0264-9381-32-13-135012}
\bibinfo{author}{\bibfnamefont{N.~J.} \bibnamefont{Cornish}} \bibnamefont{and}
  \bibinfo{author}{\bibfnamefont{T.~B.} \bibnamefont{Littenberg}},
  \bibinfo{journal}{Classical and Quantum Gravity}
  \textbf{\bibinfo{volume}{32}}, \bibinfo{pages}{135012}
  (\bibinfo{year}{2015}),
  \urlprefix\url{http://stacks.iop.org/0264-9381/32/i=13/a=135012}.

\bibitem[{\citenamefont{Abbott
  et~al.}(2018{\natexlab{a}})}]{TheLIGOScientific:2018pe}
\bibinfo{author}{\bibfnamefont{B.~P.} \bibnamefont{Abbott}}
  \bibnamefont{et~al.} (\bibinfo{collaboration}{Virgo, LIGO Scientific})
  (\bibinfo{year}{2018}{\natexlab{a}}), \eprint{1805.11579}.

\bibitem[{\citenamefont{Abbott
  et~al.}(2018{\natexlab{b}})}]{TheLIGOScientific:2018eos}
\bibinfo{author}{\bibfnamefont{B.~P.} \bibnamefont{Abbott}}
  \bibnamefont{et~al.} (\bibinfo{collaboration}{Virgo, LIGO Scientific})
  (\bibinfo{year}{2018}{\natexlab{b}}), \eprint{1805.11581}.

\bibitem[{\citenamefont{Veitch et~al.}(2015)}]{Veitch:2014wba}
\bibinfo{author}{\bibfnamefont{J.}~\bibnamefont{Veitch}} \bibnamefont{et~al.},
  \bibinfo{journal}{Phys. Rev.} \textbf{\bibinfo{volume}{D91}},
  \bibinfo{pages}{042003} (\bibinfo{year}{2015}), \eprint{1409.7215}.

\bibitem[{\citenamefont{Flanagan and Hinderer}(2008)}]{flanagan:021502}
\bibinfo{author}{\bibfnamefont{{\'{E}anna \'{E}.}.}~\bibnamefont{Flanagan}}
  \bibnamefont{and} \bibinfo{author}{\bibfnamefont{T.}~\bibnamefont{Hinderer}},
  \bibinfo{journal}{\prd} \textbf{\bibinfo{volume}{77}}, \bibinfo{eid}{021502}
  (pages~\bibinfo{numpages}{5}) (\bibinfo{year}{2008}),
  \urlprefix\url{http://link.aps.org/abstract/PRD/v77/e021502}.

\bibitem[{\citenamefont{{Hinderer}}(2008)}]{2008ApJ...677.1216H}
\bibinfo{author}{\bibfnamefont{T.}~\bibnamefont{{Hinderer}}},
  \bibinfo{journal}{\apj} \textbf{\bibinfo{volume}{677}}
  (\bibinfo{year}{2008}).

\bibitem[{\citenamefont{{Hinderer} et~al.}(2010)\citenamefont{{Hinderer},
  {Lackey}, {Lang}, and {Read}}}]{2010PhRvD..81l3016H}
\bibinfo{author}{\bibfnamefont{T.}~\bibnamefont{{Hinderer}}},
  \bibinfo{author}{\bibfnamefont{B.~D.} \bibnamefont{{Lackey}}},
  \bibinfo{author}{\bibfnamefont{R.~N.} \bibnamefont{{Lang}}},
  \bibnamefont{and} \bibinfo{author}{\bibfnamefont{J.~S.}
  \bibnamefont{{Read}}}, \bibinfo{journal}{\prd} \textbf{\bibinfo{volume}{81}}
  (\bibinfo{year}{2010}).

\bibitem[{\citenamefont{{Read} et~al.}(2009)\citenamefont{{Read}, {Markakis},
  {Shibata}, {Ury{\=u}}, {Creighton}, and {Friedman}}}]{Read:2009}
\bibinfo{author}{\bibfnamefont{J.~S.} \bibnamefont{{Read}}},
  \bibinfo{author}{\bibfnamefont{C.}~\bibnamefont{{Markakis}}},
  \bibinfo{author}{\bibfnamefont{M.}~\bibnamefont{{Shibata}}},
  \bibinfo{author}{\bibfnamefont{K.}~\bibnamefont{{Ury{\=u}}}},
  \bibinfo{author}{\bibfnamefont{J.~D.~E.} \bibnamefont{{Creighton}}},
  \bibnamefont{and} \bibinfo{author}{\bibfnamefont{J.~L.}
  \bibnamefont{{Friedman}}}, \bibinfo{journal}{\prd}
  \textbf{\bibinfo{volume}{79}}, \bibinfo{pages}{124033}
  (\bibinfo{year}{2009}), \eprint{arXiv:0901.3258}.

\bibitem[{\citenamefont{{Finn} and {Chernoff}}(1993)}]{1993PhRvD..47.2198F}
\bibinfo{author}{\bibfnamefont{L.~S.} \bibnamefont{{Finn}}} \bibnamefont{and}
  \bibinfo{author}{\bibfnamefont{D.~F.} \bibnamefont{{Chernoff}}},
  \bibinfo{journal}{\prd} \textbf{\bibinfo{volume}{47}}, \bibinfo{pages}{2198}
  (\bibinfo{year}{1993}).

\bibitem[{\citenamefont{{Flanagan} and {Hughes}}(1998)}]{1998PhRvD..57.4535F}
\bibinfo{author}{\bibfnamefont{{\'E}.~{\'E}.} \bibnamefont{{Flanagan}}}
  \bibnamefont{and} \bibinfo{author}{\bibfnamefont{S.~A.}
  \bibnamefont{{Hughes}}}, \bibinfo{journal}{\prd}
  \textbf{\bibinfo{volume}{57}}, \bibinfo{pages}{4535} (\bibinfo{year}{1998}).

\bibitem[{\citenamefont{{Allen} et~al.}(2012)\citenamefont{{Allen}, {Anderson},
  {Brady}, {Brown}, and {Creighton}}}]{2012PhRvD..85l2006A}
\bibinfo{author}{\bibfnamefont{B.}~\bibnamefont{{Allen}}},
  \bibinfo{author}{\bibfnamefont{W.~G.} \bibnamefont{{Anderson}}},
  \bibinfo{author}{\bibfnamefont{P.~R.} \bibnamefont{{Brady}}},
  \bibinfo{author}{\bibfnamefont{D.~A.} \bibnamefont{{Brown}}},
  \bibnamefont{and} \bibinfo{author}{\bibfnamefont{J.~D.~E.}
  \bibnamefont{{Creighton}}}, \bibinfo{journal}{\prd}
  \textbf{\bibinfo{volume}{85}} (\bibinfo{year}{2012}).

\bibitem[{\citenamefont{Sathyaprakash and Schutz}(2009)}]{Sathyaprakash:2009xs}
\bibinfo{author}{\bibfnamefont{B.~S.} \bibnamefont{Sathyaprakash}}
  \bibnamefont{and} \bibinfo{author}{\bibfnamefont{B.~F.}
  \bibnamefont{Schutz}}, \bibinfo{journal}{Living Rev. Rel.}
  \textbf{\bibinfo{volume}{12}}, \bibinfo{pages}{2} (\bibinfo{year}{2009}),
  \eprint{0903.0338}.

\bibitem[{\citenamefont{Powell}(2018)}]{Powell:2018csz}
\bibinfo{author}{\bibfnamefont{J.}~\bibnamefont{Powell}},
  \bibinfo{journal}{Class. Quant. Grav.} \textbf{\bibinfo{volume}{35}},
  \bibinfo{pages}{155017} (\bibinfo{year}{2018}), \eprint{1803.11346}.

\bibitem[{\citenamefont{{McIver} et~al.}(2018)\citenamefont{{McIver},
  {Massinger}, {Davis}, {Nuttall}, and {Smith}}}]{McIver2018}
\bibinfo{author}{\bibfnamefont{J.}~\bibnamefont{{McIver}}},
  \bibinfo{author}{\bibfnamefont{T.~J.} \bibnamefont{{Massinger}}},
  \bibinfo{author}{\bibfnamefont{D.}~\bibnamefont{{Davis}}},
  \bibinfo{author}{\bibfnamefont{L.~K.} \bibnamefont{{Nuttall}}},
  \bibnamefont{and} \bibinfo{author}{\bibfnamefont{R.}~\bibnamefont{{Smith}}}
  (\bibinfo{year}{2018}).

\bibitem[{\citenamefont{Abbott et~al.}(2016)}]{2016DetChar.150914}
\bibinfo{author}{\bibfnamefont{B.~P.} \bibnamefont{Abbott}}
  \bibnamefont{et~al.}, \bibinfo{journal}{CQG} \textbf{\bibinfo{volume}{33}}
  (\bibinfo{year}{2016}), \eprint{1602.03844}.

\bibitem[{\citenamefont{Veitch and Vecchio}(2010)}]{Veitch:2009hd}
\bibinfo{author}{\bibfnamefont{J.}~\bibnamefont{Veitch}} \bibnamefont{and}
  \bibinfo{author}{\bibfnamefont{A.}~\bibnamefont{Vecchio}},
  \bibinfo{journal}{Phys. Rev.} \textbf{\bibinfo{volume}{D81}},
  \bibinfo{pages}{062003} (\bibinfo{year}{2010}), \eprint{0911.3820}.

\bibitem[{\citenamefont{{Kanner} et~al.}(2015)\citenamefont{{Kanner},
  {Littenberg}, {Cornish}, {Millhouse}, {Xhakaj}, {Salemi}, {Drago},
  {Vedovato}, and {Klimenko}}}]{Kanner2015}
\bibinfo{author}{\bibfnamefont{J.}~\bibnamefont{{Kanner}}},
  \bibinfo{author}{\bibfnamefont{T.}~\bibnamefont{{Littenberg}}},
  \bibinfo{author}{\bibfnamefont{N.}~\bibnamefont{{Cornish}}},
  \bibinfo{author}{\bibfnamefont{M.}~\bibnamefont{{Millhouse}}},
  \bibinfo{author}{\bibfnamefont{E.}~\bibnamefont{{Xhakaj}}},
  \bibinfo{author}{\bibfnamefont{F.}~\bibnamefont{{Salemi}}},
  \bibinfo{author}{\bibfnamefont{M.}~\bibnamefont{{Drago}}},
  \bibinfo{author}{\bibfnamefont{G.}~\bibnamefont{{Vedovato}}},
  \bibnamefont{and}
  \bibinfo{author}{\bibfnamefont{S.}~\bibnamefont{{Klimenko}}},
  \bibinfo{journal}{\prd} \textbf{\bibinfo{volume}{93}},
  \bibinfo{pages}{022002} (\bibinfo{year}{2015}).

\bibitem[{\citenamefont{{Biswas} et~al.}(2013)\citenamefont{{Biswas},
  {Blackburn}, {Cao}, {Essick}, {Hodge}, {Katsavounidis}, {Kim}, {Kim}, {Le
  Bigot}, {Lee} et~al.}}]{2013PhRvD..88f2003B}
\bibinfo{author}{\bibfnamefont{R.}~\bibnamefont{{Biswas}}},
  \bibinfo{author}{\bibfnamefont{L.}~\bibnamefont{{Blackburn}}},
  \bibinfo{author}{\bibfnamefont{J.}~\bibnamefont{{Cao}}},
  \bibinfo{author}{\bibfnamefont{R.}~\bibnamefont{{Essick}}},
  \bibinfo{author}{\bibfnamefont{K.~A.} \bibnamefont{{Hodge}}},
  \bibinfo{author}{\bibfnamefont{E.}~\bibnamefont{{Katsavounidis}}},
  \bibinfo{author}{\bibfnamefont{K.}~\bibnamefont{{Kim}}},
  \bibinfo{author}{\bibfnamefont{Y.-M.} \bibnamefont{{Kim}}},
  \bibinfo{author}{\bibfnamefont{E.-O.} \bibnamefont{{Le Bigot}}},
  \bibinfo{author}{\bibfnamefont{C.-H.} \bibnamefont{{Lee}}},
  \bibnamefont{et~al.}, \bibinfo{journal}{\prd} \textbf{\bibinfo{volume}{88}},
  \bibinfo{eid}{062003} (\bibinfo{year}{2013}), \eprint{1303.6984}.

\bibitem[{\citenamefont{{Essick} et~al.}(2013)\citenamefont{{Essick},
  {Blackburn}, and {Katsavounidis}}}]{2013CQGra..30o5010E}
\bibinfo{author}{\bibfnamefont{R.}~\bibnamefont{{Essick}}},
  \bibinfo{author}{\bibfnamefont{L.}~\bibnamefont{{Blackburn}}},
  \bibnamefont{and}
  \bibinfo{author}{\bibfnamefont{E.}~\bibnamefont{{Katsavounidis}}},
  \bibinfo{journal}{Classical and Quantum Gravity}
  \textbf{\bibinfo{volume}{30}}, \bibinfo{eid}{155010} (\bibinfo{year}{2013}),
  \eprint{1303.7159}.

\bibitem[{\citenamefont{Littenberg and Cornish}(2015)}]{Littenberg:2014oda}
\bibinfo{author}{\bibfnamefont{T.~B.} \bibnamefont{Littenberg}}
  \bibnamefont{and} \bibinfo{author}{\bibfnamefont{N.~J.}
  \bibnamefont{Cornish}}, \bibinfo{journal}{Phys. Rev.}
  \textbf{\bibinfo{volume}{D91}}, \bibinfo{pages}{084034}
  (\bibinfo{year}{2015}), \eprint{1410.3852}.

\bibitem[{\citenamefont{Farr et~al.}(2015)\citenamefont{Farr, Farr, and
  Littenberg}}]{SplineCalMarg-T1400682}
\bibinfo{author}{\bibfnamefont{W.~M.} \bibnamefont{Farr}},
  \bibinfo{author}{\bibfnamefont{B.}~\bibnamefont{Farr}}, \bibnamefont{and}
  \bibinfo{author}{\bibfnamefont{T.}~\bibnamefont{Littenberg}},
  \bibinfo{type}{Tech. Rep.} \bibinfo{number}{{LIGO}-T1400682},
  \bibinfo{institution}{{LIGO} Project} (\bibinfo{year}{2015}),
  \urlprefix\url{https://dcc.ligo.org/LIGO-T1400682/public}.

\bibitem[{\citenamefont{Racine}(2008)}]{Racine:2008qv}
\bibinfo{author}{\bibfnamefont{E.}~\bibnamefont{Racine}},
  \bibinfo{journal}{Phys. Rev.} \textbf{\bibinfo{volume}{D78}},
  \bibinfo{pages}{044021} (\bibinfo{year}{2008}), \eprint{0803.1820}.

\bibitem[{\citenamefont{{Cutler} and {Flanagan}}(1994)}]{1994PhRvD..49.2658C}
\bibinfo{author}{\bibfnamefont{C.}~\bibnamefont{{Cutler}}} \bibnamefont{and}
  \bibinfo{author}{\bibfnamefont{{\'E}.~E.} \bibnamefont{{Flanagan}}},
  \bibinfo{journal}{\prd} \textbf{\bibinfo{volume}{49}}, \bibinfo{pages}{2658}
  (\bibinfo{year}{1994}).

\bibitem[{\citenamefont{Hannam et~al.}(2014)\citenamefont{Hannam, Schmidt,
  Boh{\'e}, Haegel, Husa, Ohme, Pratten, and P{\"u}rrer}}]{Hannam:2013oca}
\bibinfo{author}{\bibfnamefont{M.}~\bibnamefont{Hannam}},
  \bibinfo{author}{\bibfnamefont{P.}~\bibnamefont{Schmidt}},
  \bibinfo{author}{\bibfnamefont{A.}~\bibnamefont{Boh{\'e}}},
  \bibinfo{author}{\bibfnamefont{L.}~\bibnamefont{Haegel}},
  \bibinfo{author}{\bibfnamefont{S.}~\bibnamefont{Husa}},
  \bibinfo{author}{\bibfnamefont{F.}~\bibnamefont{Ohme}},
  \bibinfo{author}{\bibfnamefont{G.}~\bibnamefont{Pratten}}, \bibnamefont{and}
  \bibinfo{author}{\bibfnamefont{M.}~\bibnamefont{P{\"u}rrer}},
  \bibinfo{journal}{Phys. Rev. Lett.} \textbf{\bibinfo{volume}{113}},
  \bibinfo{pages}{151101} (\bibinfo{year}{2014}), \eprint{1308.3271}.

\bibitem[{\citenamefont{Schmidt et~al.}(2015)\citenamefont{Schmidt, Ohme, and
  Hannam}}]{Schmidt:2014iyl}
\bibinfo{author}{\bibfnamefont{P.}~\bibnamefont{Schmidt}},
  \bibinfo{author}{\bibfnamefont{F.}~\bibnamefont{Ohme}}, \bibnamefont{and}
  \bibinfo{author}{\bibfnamefont{M.}~\bibnamefont{Hannam}},
  \bibinfo{journal}{Phys. Rev. D} \textbf{\bibinfo{volume}{91}},
  \bibinfo{pages}{024043} (\bibinfo{year}{2015}), \eprint{1408.1810}.

\bibitem[{\citenamefont{Husa et~al.}(2016)\citenamefont{Husa, Khan, Hannam,
  P{\"u}rrer, Ohme, Forteza, and Boh{\'e}}}]{Husa:2015iqa}
\bibinfo{author}{\bibfnamefont{S.}~\bibnamefont{Husa}},
  \bibinfo{author}{\bibfnamefont{S.}~\bibnamefont{Khan}},
  \bibinfo{author}{\bibfnamefont{M.}~\bibnamefont{Hannam}},
  \bibinfo{author}{\bibfnamefont{M.}~\bibnamefont{P{\"u}rrer}},
  \bibinfo{author}{\bibfnamefont{F.}~\bibnamefont{Ohme}},
  \bibinfo{author}{\bibfnamefont{X.~J.} \bibnamefont{Forteza}},
  \bibnamefont{and} \bibinfo{author}{\bibfnamefont{A.}~\bibnamefont{Boh{\'e}}},
  \bibinfo{journal}{Phys. Rev. D} \textbf{\bibinfo{volume}{93}},
  \bibinfo{pages}{044006} (\bibinfo{year}{2016}), \eprint{1508.07250}.

\bibitem[{\citenamefont{Khan et~al.}(2016)\citenamefont{Khan, Husa, Hannam,
  Ohme, P{\"{u}}rrer, Forteza, and Boh{\'e}}}]{Khan:2015jqa}
\bibinfo{author}{\bibfnamefont{S.}~\bibnamefont{Khan}},
  \bibinfo{author}{\bibfnamefont{S.}~\bibnamefont{Husa}},
  \bibinfo{author}{\bibfnamefont{M.}~\bibnamefont{Hannam}},
  \bibinfo{author}{\bibfnamefont{F.}~\bibnamefont{Ohme}},
  \bibinfo{author}{\bibfnamefont{M.}~\bibnamefont{P{\"{u}}rrer}},
  \bibinfo{author}{\bibfnamefont{X.~J.} \bibnamefont{Forteza}},
  \bibnamefont{and} \bibinfo{author}{\bibfnamefont{A.}~\bibnamefont{Boh{\'e}}},
  \bibinfo{journal}{Phys. Rev. D} \textbf{\bibinfo{volume}{93}},
  \bibinfo{pages}{044007} (\bibinfo{year}{2016}), \eprint{1508.07253}.

\bibitem[{\citenamefont{{Damour} et~al.}(2002)\citenamefont{{Damour}, {Iyer},
  and {Sathyaprakash}}}]{2002PhRvD..66b7502D}
\bibinfo{author}{\bibfnamefont{T.}~\bibnamefont{{Damour}}},
  \bibinfo{author}{\bibfnamefont{B.~R.} \bibnamefont{{Iyer}}},
  \bibnamefont{and} \bibinfo{author}{\bibfnamefont{B.~S.}
  \bibnamefont{{Sathyaprakash}}}, \bibinfo{journal}{\prd}
  \textbf{\bibinfo{volume}{66}}, \bibinfo{eid}{027502} (\bibinfo{year}{2002}).

\bibitem[{\citenamefont{{Vines} et~al.}(2011)\citenamefont{{Vines}, {Flanagan},
  and {Hinderer}}}]{2011PhRvD..83h4051V}
\bibinfo{author}{\bibfnamefont{J.}~\bibnamefont{{Vines}}},
  \bibinfo{author}{\bibfnamefont{{\'E}.~{\'E}.} \bibnamefont{{Flanagan}}},
  \bibnamefont{and}
  \bibinfo{author}{\bibfnamefont{T.}~\bibnamefont{{Hinderer}}},
  \bibinfo{journal}{\prd} \textbf{\bibinfo{volume}{83}}, \bibinfo{eid}{084051}
  (\bibinfo{year}{2011}).

\bibitem[{\citenamefont{Dietrich et~al.}(2018)}]{Dietrich:2018uni}
\bibinfo{author}{\bibfnamefont{T.}~\bibnamefont{Dietrich}} \bibnamefont{et~al.}
  (\bibinfo{year}{2018}), \eprint{1804.02235}.

\bibitem[{\citenamefont{Littenberg et~al.}(2013)\citenamefont{Littenberg,
  Baker, Buonanno, and Kelly}}]{Littenberg:2012uj}
\bibinfo{author}{\bibfnamefont{T.~B.} \bibnamefont{Littenberg}},
  \bibinfo{author}{\bibfnamefont{J.~G.} \bibnamefont{Baker}},
  \bibinfo{author}{\bibfnamefont{A.}~\bibnamefont{Buonanno}}, \bibnamefont{and}
  \bibinfo{author}{\bibfnamefont{B.~J.} \bibnamefont{Kelly}},
  \bibinfo{journal}{Phys. Rev.} \textbf{\bibinfo{volume}{D87}},
  \bibinfo{pages}{104003} (\bibinfo{year}{2013}), \eprint{1210.0893}.

\bibitem[{\citenamefont{{Rodriguez} et~al.}(2014)\citenamefont{{Rodriguez},
  {Farr}, {Raymond}, {Farr}, {Littenberg}, {Fazi}, and
  {Kalogera}}}]{2014ApJ...784..119R}
\bibinfo{author}{\bibfnamefont{C.~L.} \bibnamefont{{Rodriguez}}},
  \bibinfo{author}{\bibfnamefont{B.}~\bibnamefont{{Farr}}},
  \bibinfo{author}{\bibfnamefont{V.}~\bibnamefont{{Raymond}}},
  \bibinfo{author}{\bibfnamefont{W.~M.} \bibnamefont{{Farr}}},
  \bibinfo{author}{\bibfnamefont{T.~B.} \bibnamefont{{Littenberg}}},
  \bibinfo{author}{\bibfnamefont{D.}~\bibnamefont{{Fazi}}}, \bibnamefont{and}
  \bibinfo{author}{\bibfnamefont{V.}~\bibnamefont{{Kalogera}}},
  \bibinfo{journal}{\apj} \textbf{\bibinfo{volume}{784}}, \bibinfo{eid}{119}
  (\bibinfo{year}{2014}), \eprint{1309.3273}.

\bibitem[{\citenamefont{Vitale and Chen}(2018)}]{PhysRevLett.121.021303}
\bibinfo{author}{\bibfnamefont{S.}~\bibnamefont{Vitale}} \bibnamefont{and}
  \bibinfo{author}{\bibfnamefont{H.-Y.} \bibnamefont{Chen}},
  \bibinfo{journal}{Phys. Rev. Lett.} \textbf{\bibinfo{volume}{121}},
  \bibinfo{pages}{021303} (\bibinfo{year}{2018}),
  \urlprefix\url{https://link.aps.org/doi/10.1103/PhysRevLett.121.021303}.

\bibitem[{\citenamefont{Nissanke et~al.}(2010)\citenamefont{Nissanke, Holz,
  Hughes, Dalal, and Sievers}}]{Nissanke:2009kt}
\bibinfo{author}{\bibfnamefont{S.}~\bibnamefont{Nissanke}},
  \bibinfo{author}{\bibfnamefont{D.~E.} \bibnamefont{Holz}},
  \bibinfo{author}{\bibfnamefont{S.~A.} \bibnamefont{Hughes}},
  \bibinfo{author}{\bibfnamefont{N.}~\bibnamefont{Dalal}}, \bibnamefont{and}
  \bibinfo{author}{\bibfnamefont{J.~L.} \bibnamefont{Sievers}},
  \bibinfo{journal}{Astrophys. J.} \textbf{\bibinfo{volume}{725}},
  \bibinfo{pages}{496} (\bibinfo{year}{2010}), \eprint{0904.1017}.

\bibitem[{\citenamefont{Vallisneri}(2011)}]{Vallisneri:2011ts}
\bibinfo{author}{\bibfnamefont{M.}~\bibnamefont{Vallisneri}},
  \bibinfo{journal}{Phys. Rev. Lett.} \textbf{\bibinfo{volume}{107}},
  \bibinfo{pages}{191104} (\bibinfo{year}{2011}), \eprint{1108.1158}.

\bibitem[{\citenamefont{{Abbott} et~al.}(2016)\citenamefont{{Abbott}, {Abbott},
  {Abbott}, {Abernathy}, {Acernese}, {Ackley}, {Adams}, {Adams}, {Addesso},
  {Adhikari} et~al.}}]{2016LRR....19....1A}
\bibinfo{author}{\bibfnamefont{B.~P.} \bibnamefont{{Abbott}}},
  \bibinfo{author}{\bibfnamefont{R.}~\bibnamefont{{Abbott}}},
  \bibinfo{author}{\bibfnamefont{T.~D.} \bibnamefont{{Abbott}}},
  \bibinfo{author}{\bibfnamefont{M.~R.} \bibnamefont{{Abernathy}}},
  \bibinfo{author}{\bibfnamefont{F.}~\bibnamefont{{Acernese}}},
  \bibinfo{author}{\bibfnamefont{K.}~\bibnamefont{{Ackley}}},
  \bibinfo{author}{\bibfnamefont{C.}~\bibnamefont{{Adams}}},
  \bibinfo{author}{\bibfnamefont{T.}~\bibnamefont{{Adams}}},
  \bibinfo{author}{\bibfnamefont{P.}~\bibnamefont{{Addesso}}},
  \bibinfo{author}{\bibfnamefont{R.~X.} \bibnamefont{{Adhikari}}},
  \bibnamefont{et~al.}, \bibinfo{journal}{Living Reviews in Relativity}
  \textbf{\bibinfo{volume}{19}}, \bibinfo{eid}{1} (\bibinfo{year}{2016}).

\end{thebibliography}

\end{document}